\documentclass[prd,preprint,superscriptaddress,preprintnumbers,eqsecnum,showpacs,nofootinbib,nobibnotes]{revtex4}
\usepackage{amsfonts,amsmath,amssymb,bm,natbib}
\usepackage{graphicx} 
\usepackage{pstricks}

\newcommand{\be}{\begin{equation}}
\newcommand{\bea}{\begin{eqnarray}}
\newcommand{\ee}{\end{equation}}
\newcommand{\eea}{\end{eqnarray}}
\def\s#1{{\scriptscriptstyle #1}}

\def\1eq#1{Eq.~(\ref{#1})}

\def\2eqs#1#2{Eqs.~(\ref{#1}) and~(\ref{#2})}
\def\3eqs#1#2#3{Eqs.~(\ref{#1}),~(\ref{#2}) and~(\ref{#3})}
\def\noeq#1{(\ref{#1})}

\def\fig#1{Fig.~\ref{#1}}

\def\chic#1{{\scriptscriptstyle #1}}

\def\diff#1{{\rm d}^#1\,}

\def\ie{{\it i.e.}, }
\def\eg{{\it e.g.}, }

\def\Nf{N_{\!f}}

\def\gtree{\Gamma^{(0)}}

\def\gfullb{\widetilde{\Gamma}}

\def\bcj{J}

\def\NV{\gfullb'}

\def\NP{\widetilde{V}}

\def\mrgi{\overline m}

\def\s#1{{\scriptscriptstyle #1}}

\def\n#1{({\it #1}\,)}

\def\gA{g^2 C_A}

\def\Y{Y}

\begin{document}

\title{Renormalization group analysis of the gluon mass equation}

\author{A.~C. Aguilar}
\affiliation{University of Campinas - UNICAMP, Institute of Physics ``Gleb Wataghin'',
13083-859 Campinas, SP, Brazil}
\author{D. Binosi}
\affiliation{European Centre for Theoretical Studies in Nuclear
Physics and Related Areas (ECT*) and Fondazione Bruno Kessler, \\Villa Tambosi, Strada delle
Tabarelle 286, 
I-38123 Villazzano (TN)  Italy}
\author{J. Papavassiliou}
\affiliation{\mbox{Department of Theoretical Physics and IFIC, 
University of Valencia and CSIC},
E-46100, Valencia, Spain}

\begin{abstract}

In the present work 
we carry out a
systematic study of the renormalization properties  of the 
integral equation that determines the momentum evolution of
the effective gluon  mass. 
A detailed, all-order analysis of the complete kernel 
appearing in this particular equation 
reveals that the renormalization procedure   
may be accomplished through the sole use 
of ingredients known from the 
standard  perturbative treatment of the theory, 
with no additional assumptions. 
However, the subtle interplay of terms operating at the level of the exact equation gets distorted by the approximations usually employed when evaluating the aforementioned kernel. This fact is reflected in the form of the obtained solutions, whose deviations from the correct behavior are best quantified by resorting to appropriately defined renormalization-group invariant quantities.
This analysis, in turn, provides a solid guiding principle 
for improving the form of the kernel, 
and furnishes a well-defined criterion for 
discriminating between various possibilities. 
Certain renormalization-group inspired Ans\"atze for the kernel 
are then proposed, and their numerical implications 
are explored in detail. One of the solutions  
obtained fulfills the theoretical expectations to a high degree of accuracy, 
yielding a gluon mass that is positive-definite 
throughout the entire range of physical momenta, and  
displays in the ultraviolet the  so-called ``power-law''
running, in agreement with standard arguments based 
on the  operator product expansion. 
Some of the technical difficulties thwarting a 
more rigorous  determination of the kernel are discussed, 
and possible future directions are briefly mentioned.

\end{abstract}


\maketitle

\section{Introduction}

A recent development in 
the ongoing study of 
the basic QCD Green's functions  
within the nonperturbative framework of the Schwinger-Dyson equations (SDEs)~\cite{Roberts:1994dr,
Curtis:1993py,Kizilersu:2009kg,Boucaud:2008ji,Boucaud:2008ky,Alkofer:2000wg,
Fischer:2006ub,Szczepaniak:2003ve,Szczepaniak:2001rg,Aguilar:2006gr,Aguilar:2008xm,Pennington:2011xs,
Huber:2012kd,Campagnari:2011bk,Szczepaniak:2011bs,Pawlowski:2005xe,Pawlowski:2003hq,Popovici:2011pw} is the 
derivation of the particular integral equation that governs the momentum evolution of the effective 
gluon mass~\cite{Cornwall:1981zr,Aguilar:2011ux,Binosi:2012sj,Aguilar:2013hoa}. As has been argued in a series of works~\cite{Nair:2013kva,Philipsen:2001ip,Aguilar:2002tc,Aguilar:2001zy}, 
the generation of such a mass offers a natural and self-consistent explanation for the  
infrared finiteness of the (Landau gauge) gluon propagator and ghost 
dressing function~\cite{Cornwall:1981zr, Aguilar:2008xm,Aguilar:2011ux,Aguilar:2009nf,Dudal:2008sp}, established in large-volume lattice simulations, both in  
$SU(2)$~\cite{Cucchieri:2007md} and in $SU(3)$~\cite{Bowman:2007du,Bogolubsky:2009dc,Oliveira:2009nn,Ayala:2012pb}. 

The systematic scrutiny of this equation could eventually 
place the  gluon mass generation 
on an equal conceptual footing 
as the more familiar phenomenon of constituent quark mass generation~\cite{Roberts:1994dr,Fischer:2003rp,
Papavassiliou:1991hx,Williams:1989tv,Aguilar:2010cn,August:2013jia}.
In order to reach an equivalent level of 
understanding, however, several theoretical 
tasks need be carried out. In particular, 
one of the main unresolved issues in this context 
is the proper renormalization of this homogeneous integral equation.
The renormalization procedure, in turn, may impose crucial restrictions on the form of its kernel,
which, even though is formally known, for all practical 
purposes must undergo approximations and modelling~\cite{Binosi:2012sj}.

In the present work we study in detail 
the general renormalization procedure, and, most importantly, 
the properties of the 
mass equation, and its corresponding solutions, under the 
renormalization group (RG). 
This is a rather technical endeavor, whose main field-theoretic points 
may be summarized as follows.

To begin with, 
it is important to recognize
that the renormalization of the mass equation, as well as the gluon mass itself, is accomplished 
entirely by means of the same renormalization constants familiar from the 
perturbative treatment of Yang-Mills theories, 
namely those associated with the gluon and ghost propagators, 
and the various interaction vertices~\cite{Bjorken:1979dk}.  
The deeper field-theoretic reasons for this fact may be traced back to the intricate  
dynamical mechanism that generates this effective mass; specifically, 
the formation of non-perturbative  massless bound states~\cite{Jackiw:1973tr,Jackiw:1973ha,Cornwall:1973ts,Eichten:1974et,Poggio:1974qs}, which 
act as would-be Goldstone bosons, trigger the well-known Schwinger mechanism~\cite{Schwinger:1962tn,Schwinger:1962tp},  
without ever modifying the original  Lagrangian. In addition, a crucial identity   
enforces the total annihilation of any potential quadratic divergence, 
related to seagull-type integrals~\cite{Aguilar:2009ke}.
As a result, no bare gluon mass needs be introduced at any stage; this is 
absolutely essential,  
since a term of the type $m^2_0 A^2_{\mu}$ is 
forbidden by the local gauge invariance of the Yang-Mills Lagrangian~\cite{Cornwall:1981zr,Aguilar:2006gr}.

Furthermore, 
it is clear that the correct implementation of 
the aforementioned renormalization procedure relies crucially on the 
precise properties of the kernel of the mass equation under the RG.
If the kernel is treated at the formal level, 
these properties are 
automatically enforced, as a direct consequence of the 
corresponding RG properties of the basic ingredients that build it up.
However, the kernel is expressed in terms of a complicated diagrammatic expansion, 
which, for all practical purposes, must be truncated, and further simplified or approximated~\cite{Binosi:2012sj}. 
As a result, the exact RG properties of the kernel may be 
compromised; this flaw, in turn, will make its way into the 
solutions obtained from the corresponding mass equation. 
Thus, depending on the quality of the approximations employed for the kernel, 
the corresponding gluon masses will encode their formal 
RG properties with variable degrees of accuracy.

The quantitative study of the situation described above may be best 
accomplished by using  
RG invariant (RGI) quantities, which, by construction, 
maintain the same form before and after renormalization, 
and are independent 
of the value of the renormalization point $\mu$, used to implement the  
various subtractions~\cite{Cornwall:1981zr,Aguilar:2009nf}.
In particular, an RGI gluon mass 
may be defined, and then subsequently constructed from  
the solutions of the mass equation, for any given Ansatz for the 
kernel, and for several different values of $\mu$.
Then, the amount by which the resulting quantity 
departs from the  
perfect $\mu$-independence can serve as a discriminant of the 
various possible  Ans\"atze for the kernel.

The article is organized as follows. 
In Section~\ref{RandM} we introduce the relevant notation, define  
the basic renormalization constants, 
and discuss in detail the particularities of the 
gluon mass and its renormalization.
In Section~\ref{RGprop} we carry out 
the full renormalization of the gluon mass equation, and 
explore its properties under the RG.
Then, in Section ~\ref{tenstr} we study the tensorial structure of the 
main unknown ingredient that composes the kernel, 
and determine how its various 
form factors affect the infrared behavior of the mass equation.
In Section~\ref{RGorig} we outline the procedure 
for estimating the  
discrepancies from the correct RG behavior induced 
by the various approximations to the kernel. 
This procedure is then applied to the 
original version of the mass equation, and considerable 
deviations are found. 
In Section~\ref{RGimp} we present two RG-inspired improvements of the kernel, 
which, a priori, seem to capture more faithfully its formal   
RG properties, 
and determine the corresponding departures of the new solutions from the ideal $\mu$-independence.  
This study reveals a significant improvement, in accordance with the initial expectation.
The asymptotic behavior of one of these ``improved''solutions is further analyzed, suggesting a 
possible connection with general arguments originating from the 
operator product expansion (OPE)~\cite{Wilson:1969zs,Shifman:1978bx,Shifman:1978by}.
Finally, in Section~\ref{conc}  we present our discussion and conclusions.

\section{\label{RandM} Renormalization and the gluon mass}

In this section we set up the notation, and introduce the field theoretic  
relations and concepts necessary for carrying out the renormalization of the 
gluon mass equation, and for exploring its properties under the RG.

\subsection{General renormalization relations}

In the Landau gauge, the full gluon propagator (quenched or unquenched) 
assumes the general form 
\be
i{\Delta}_{\mu\nu}(q)=-i{\Delta}(q^2)P_{\mu\nu}(q); \qquad 
P_{\mu\nu}(q)=g_{\mu\nu}-{q_\mu q_\nu}/{q^2}.
\label{prop-def}
\ee 
At any finite order in perturbation theory, the scalar cofactor $\Delta(q^2)$ 
is conveniently parametrized in terms of the 
\textit{inverse} gluon dressing function, $J(q^2)$, 
\begin{equation}
\Delta^{-1}(q^2)=q^2 J(q^2).   
\label{massless}
\end{equation}
In addition, the full ghost propagator, $D(q^2)$, is usually parameterized 
in terms of the corresponding 
ghost dressing function, $F(q^2)$, according to  
\be
D(q^2) = \frac{F(q^2)}{q^2}.
\ee

We will now consider the combination of the pinch technique (PT)~\cite{Cornwall:1981zr,Cornwall:1989gv,Pilaftsis:1996fh, 
Binosi:2002ft,Binosi:2003rr,Binosi:2009qm}  with the background field method (BFM)~\cite{Abbott:1980hw}, 
known as the PT-BFM scheme~\cite{Aguilar:2006gr,Binosi:2007pi,Binosi:2008qk}.  Within the PT-BFM formalism,  
the natural separation of the gluonic field
into a ``quantum'' ($Q$) and a ``background'' ($B$) part, gives rise to 
an increase in the type of possible Green's functions that 
one may consider~\cite{Abbott:1980hw}. In particular, 
three types of gluon propagator make their appearance: 
\n{\it i} the conventional gluon propagator (two quantum gluons entering, $QQ$), 
denoted (as above) by $\Delta(q^2)$; \n{\it ii}
the background  gluon propagator (two background gluons entering, $BB$), 
indicated by $\widehat\Delta(q^2)$; and \n{\it iii}
the mixed  background-quantum gluon propagator (one background and 
one quantum gluons entering, $BQ$), denoted by  $\widetilde\Delta(q^2)$. 

The conversion 
between quantum and background two-point functions is achieved through the so-called 
background-quantum identities (BQIs)~\cite{Grassi:1999tp,Binosi:2002ez}.
For instance, $\widehat\Delta$ and $\Delta$, as well as their corresponding components, are related by
\bea
\widehat{O}(q^2) &=&[1+G(q^2)]^2 O(q^2);\qquad O=\Delta^{-1},\ J,\ m^2,
\nonumber\\
\widetilde{O}(q^2) &=&[1+G(q^2)] O(q^2) .
\label{BQI}
\eea

The function $G(q^2)$ represents the $g_{\mu\nu}$ component 
of a special Green's function, $\Lambda_{\mu\nu}(q)$, typical of the PT-BFM framework~\cite{Binosi:2002ft}, 
\ie $\Lambda_{\mu\nu}(q) = G(q^2)g_{\mu\nu} +  L(q^2) q_{\mu}q_{\nu}/q^2$; for 
various field-theoretic properties of the above functions, see~\cite{Aguilar:2009pp} and references therein. 
Here it should suffice to 
mention that, for practical purposes, one often uses the approximate (but rather accurate)   
relation 
\be
 1 + G(q^2) \approx F^{-1}(q^2),
\label{GFapp}
\ee
which becomes exact in the deep IR~\cite{Grassi:2004yq,Aguilar:2009nf,Aguilar:2009pp,Aguilar:2010gm}.

At any finite order in perturbation theory, 
the renormalization of the pure Yang-Mills theory  
proceeds through the standard redefinition of the bare 
fundamental fields, gluon $A_0^{a\,\mu}(x)$, and ghost $c_0^a(x)$, and the 
bare gauge coupling, $g_0$; specifically,  the corresponding renormalized quantities,
$A^{a\,\mu}_{\chic R}(x)$, $c^a_{\chic R}(x)$, and $g_{\chic R}$, 
are given by  
\be 
A^{a\,\mu}_{\chic R}(x) = Z^{-1/2}_{\chic A} A_0^{a\,\mu}(x),\qquad
c_{\chic R}^a(x) = Z^{-1/2}_{c} c_0^a(x);\qquad 
g_{\chic R} = Z_g^{-1} g_0.
\ee 
Then the associated two point functions are renormalized as 
\be
\Delta_{\chic R}(q^2) =  Z^{-1}_{\chic A} \Delta_0(q^2);\qquad
D_{\chic R}(q^2) = Z^{-1}_{c} D_0(q^2),
\label{renprop}
\ee
or, equivalently, 
\be
J_{\chic R}(q^2) =  Z_{\chic A} J_0(q^2);\qquad
F_{\chic R}(q^2) = Z^{-1}_{c} F_0(q^2).
\label{renwave}
\ee
Similarly, the renormalization constants of the three fundamental Yang-Mills vertices
(gluon-ghost, three-gluon, and four-gluon vertices) are defined as~\cite{Pascual:1984zb} 
\be
\Gamma^{\mu}_{\chic R} = Z_1 \Gamma^{\mu}_0;\qquad  
\Gamma^{\mu\alpha\beta}_{\chic R} = Z_3 \Gamma^{\mu\alpha\beta}_0;\qquad
\Gamma^{\mu\alpha\beta\nu}_{\chic R} = Z_4 \Gamma^{\mu\alpha\beta\nu}_0.
\label{renconst2}
\ee
The standard Slavnov-Taylor identities (STIs) of the theory enforce 
a set of important relations on the 
various renormalization constants~\cite{Pascual:1984zb}, namely   
\be
Z_g = Z_1 Z_{\chic A}^{-1/2} Z_c^{-1} = Z_3  Z_{\chic A}^{-3/2} = Z_4^{1/2} Z_{\chic A}^{-1},
\label{STIrel}
\ee
which will be used extensively in Section III. 

In the BFM one introduces, in addition, the wave-function renormalization constant 
${\widehat Z}_{\chic A}$, associated with the background gluon $B$. Then, 
$\widehat\Delta(q^2)$ renormalizes according to
\be
\widehat\Delta_{\chic R}(q^2) =  {\widehat Z}^{-1}_{\chic A} \widehat\Delta_0(q^2).
\ee
Due to the Abelian  Ward Identities (WIs) of the BFM, $Z_g$ and ${\widehat Z}_{\chic A}$ are related by the   
fundamental  QED-like  relation~\cite{Abbott:1980hw}
\be
Z_g =  {\widehat Z}_{\chic A}^{-1/2}. 
\label{renBFM}
\ee
Finally, the renormalization  relation for $G(q^2)$ reads~\cite{Aguilar:2009pp} 
\be
1+G_{\chic R}(q^2) = Z_{G}[1+G_0(q^2)],  
\ee
where, due to \2eqs{BQI}{renBFM}, 
$Z_{G} = {\widehat Z}^{1/2}_{\chic A} {Z}^{-1/2}_{\chic A} =  Z_g^{-1}  {Z}^{-1/2}_{\chic A}$.   

\subsection{Gluon mass renormalization} 

Nonperturbatively, the dynamical generation of an effective gluon mass 
accounts for the infrared finiteness of the (Landau gauge) 
gluon propagator, observed in a variety of large-volume lattice simulations~\cite{Cucchieri:2007md,Bowman:2007du,Bogolubsky:2009dc,Oliveira:2009nn}.
To describe this behavior, the parametrization in \1eq{massless} is modified according to 
(Minkowski space)~\cite{Aguilar:2011ux}  
\begin{equation}
\Delta^{-1}(q^2)=q^2 J(q^2)- m^2(q^2),   
\label{massive}
\end{equation}
with ${m}^2(0) \neq 0$. 
In addition, the generation of the aforementioned mass  
explains also, in a natural way, the corresponding saturation of the ghost dressing function, $F(q^2)$~\cite{Boucaud:2008ji,Aguilar:2008xm}.
Moreover, 
both $\widehat\Delta(q^2)$ and $\widetilde\Delta(q^2)$ are also infrared finite, 
and must be parametrized in a way exactly analogous to that of 
$\Delta(q^2)$ in \1eq{massive}, namely in terms of $\widehat{J}(q^2)$, $\widehat{m}^2(q^2)$, and  
$\widetilde{J}(q^2)$, $\widetilde{m}^2(q^2)$, respectively~\cite{Binosi:2012sj}. 

It is important to emphasize that 
the generation of a gluon mass does not interfere, in any way, with 
the renormalization of the theory, which proceeds exactly as before.
In particular, the following main points must be stressed:

\n{i} The Lagrangian of the Yang-Mills theory (or that of QCD) is never altered;  
the generation of the gluon mass takes place dynamically, 
without violating any of the underlying symmetries.
In particular, no bare gluon mass is introduced, 
since a term of the type $m^2_0 A^2_{\mu}$ is 
forbidden by the local gauge invariance.

However, although no such term is ab-initio introduced, the possible appearance of 
the so-called ``seagull'' divergences at later stages of the analysis 
could force its emergence.
Such divergences are 
produced by integrals of the type $\int_k \Delta(k)$, or variations thereof~\cite{Aguilar:2009ke};  
in dimensional regularization they give rise to terms of the type $m_0^{2} (1/\epsilon)$,
while, in the case of a hard  cutoff $\Lambda$, they correspond to terms proportional to $\Lambda^2$ 
(quadratic divergences). Evidently, 
their disposal would require the introduction in the original 
Lagrangian of a 
counter-term of the form $m^2_0 A^2_{\mu}$, which would be violating the basic assumptions 
stated above. Nonetheless, it turns out that, due to a set of subtle relations, particular to the 
PT-BFM framework, all such divergences are completely canceled~\cite{Aguilar:2009ke}.

\n{ii} Even though there is no ``bare gluon mass'', in the sense explained above, 
the momentum-dependent $m^2(q^2)$ undergoes renormalization, which, however, 
is not associated with 
a new renormalization constant, but is implemented by the (already existing) 
wave-function renormalization constant of the gluon, namely  $Z_{\chic A}$.
Specifically, from \1eq{massive}, and given that $\Delta^{-1}(0) = m^2(0)$,
we have that the gluon masses before and after renormalization are related by 
\be
m^2_{\chic R}(q^2) = Z_{\chic A} m^2_0(q^2).
\label{glmassren}
\ee

\n{iii}
The above renormalization condition is fully consistent with (and may be 
independently derived from) 
the general procedure that implements 
the gauge-invariant (\ie STI-preserving) generation of a gluon mass. 
Specifically, within the PT-BFM framework, 
the fully dressed vertex $BQ^2$, before mass generation,
satisfies the WI~\cite{Binosi:2009qm}
\be
q^\alpha\widetilde{\Gamma}_{\alpha\mu\nu}(q,r,p)=
p^2\bcj(p^2)P_{\mu\nu}(p)-r^2\bcj(r^2)P_{\mu\nu}(r),
\label{GWI}
\ee
and, given the first relation in \1eq{renwave}, 
the corresponding 
vertex renormalization constant, ${\widetilde Z}_3$, must obey 
\be
{\widetilde Z}_3 = Z_{\chic A}.
\label{GVWI}
\ee 
Then, for gauge-invariance to be preserved, one must  
modify this vertex according to~\cite{Aguilar:2011ux}  
\be
\NV_{\alpha\mu\nu}(q,r,p) = \left[\widetilde{\Gamma}(q,r,p) + \NP(q,r,p)\right]_{\alpha\mu\nu}, 
\label{NV}
\ee
where the special vertex $\NP(q,r,p)$ is completely longitudinal, \ie 
\be
P^{\alpha'\alpha}(q) P^{\mu'\mu}(r) P^{\nu'\nu}(p) \widetilde{V}_{\alpha\mu\nu}(q,r,p)  = 0 ,
\label{totlon}
\ee 
and 
contains massless poles of purely non-perturbative origin~\cite{Jackiw:1973tr,Jackiw:1973ha,Cornwall:1973ts,Eichten:1974et,Poggio:1974qs}, which will be ultimately responsible for  triggering  the well-known Schwinger mechanism~\cite{Schwinger:1962tn,Schwinger:1962tp}. Now, $\widetilde{\Gamma}$ and $\NV_{\alpha\mu\nu}$ must be 
renormalized by ${\widetilde Z}_3$, and so, 
\be
\NP_{\chic R}^{\alpha\mu\nu}(q,r,p)  = Z_{\chic A} \NP_0^{\alpha\mu\nu}(q,r,p).
\ee
Since, in order for the WIs to remain intact,  
$\NP_{\alpha\mu\nu}$  must satisfy 
\be
q^\alpha \NP_{\alpha\mu\nu}(q,r,p)= m^2(r^2)P_{\mu\nu}(r) - m^2(p^2)P_{\mu\nu}(p), 
\label{winp}
\ee
one finally concludes that \1eq{glmassren} must be fulfilled.

\n{iv}  We emphasize that the ``mass renormalization'' introduced above is not associated with 
a counter-term of the type $\delta m^2 = m^2_{\chic R} -  m_0^2$, as is typical in 
the case of hard boson masses, such as in scalar theories, or the electroweak sector of the 
Standard Model. 
Instead, it is akin to the renormalization that higher order Green's functions must undergo, in order 
to be made finite, even though no individual counter-term is assigned to them. 

Consider, for instance, a  scalar $\phi^4$ theory (in $d=4$), and the (one-particle irreducible)  $n$-point functions, 
$G^{(n)}(p_i)$ with $n\geq 5$. It is well-known that any such function ought to be made finite 
by means of the renormalization constants already defined for $n\leq 5$, since no counter-terms of the 
form $\phi^n$ (with $n\geq 5$) are allowed~\cite{Pascual:1984zb}. Indeed, $G^{(n)}(p_i)$ can be made finite by expressing the 
bare mass and coupling constant ($\mu_0, \lambda_0$) 
in terms of their renormalized counterparts ($\mu_{\chic R}, \lambda_{\chic R}$), 
and then 
multiplying by $Z_{\phi}^{1/2}$ for each external leg, \ie
$G^{(n)}_{\chic R}(p_i,\mu_{\chic R},\lambda_{\chic R}) = Z_{\phi}^{n/2} G^{(n)}_0(p_i,\mu_0,\lambda_0)$. 


\begin{figure}[!t]
\includegraphics[scale=0.5]{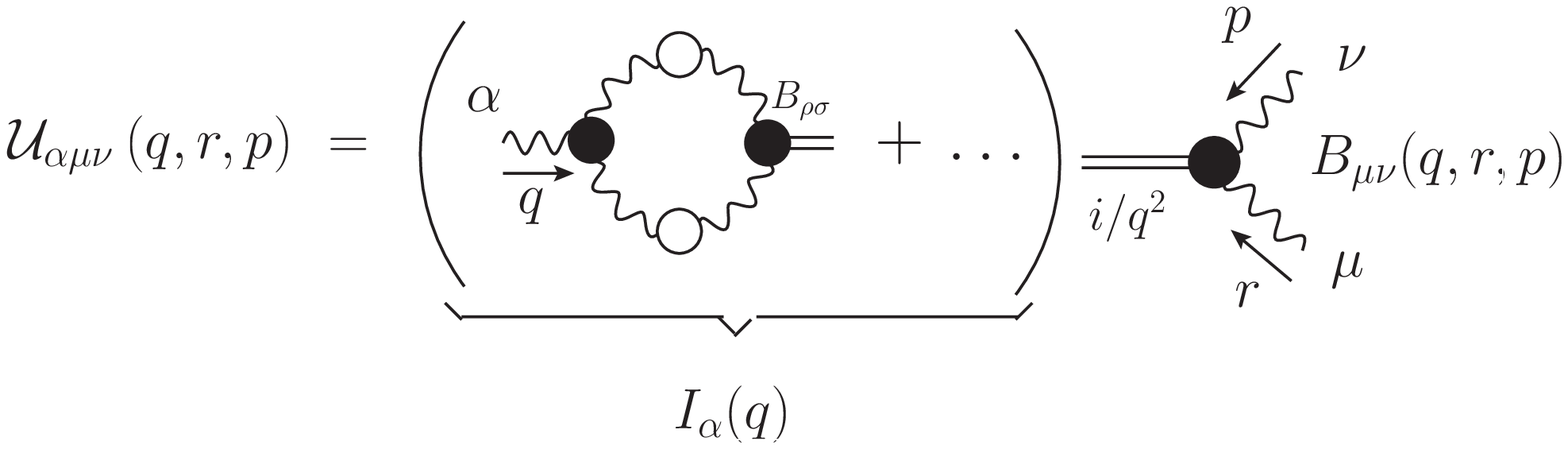}  
\caption{\label{uexpansion} The vertex $\mathcal{U_{\alpha\mu\nu}}$
is composed of three main ingredients: the transition amplitude, $I_{\alpha}(q)$, which mixes the gluon with a massless excitation, 
the propagator of the massless excitation $i/q^2$, and the massless excitation gluon vertex $B_{\mu\nu}$. 
The omitted terms are not relevant for this analysis; they can be found in~\cite{Aguilar:2011xe,Ibanez:2012zk}.}
\end{figure}


A similar situation 
arises when considering the gluon mass within 
the so-called ``massless bound-state formalism''~\cite{Aguilar:2011xe,Ibanez:2012zk},
where 
one focuses on the details of the nonperturbative formation of the pole vertex mentioned in \n{iii}. 
Specifically, the relevant vertex part, denoted by ${\cal U}_{\alpha\mu\nu}$ (see Fig.~\ref{uexpansion}), has the form 
\be
{\cal U}_{\alpha\mu\nu}(q,r,p) = I_\alpha(q)\left(\frac{i}{q^2}\right) B_{\mu\nu}(q,r,p),
\label{Uvert}
\ee
where $I_\alpha(q)$ represents the transition amplitude that mixes  a quantum gluon
with the massless excitation, $i/q^2$ corresponds to the propagator of
the massless excitation, and $B$  is an effective vertex describing the
interaction  between   the  massless  excitation and  gluons. Obviously, Lorentz invariance dictates that 
$I_\alpha(q) = q_\alpha I(q^2)$.
In addition~\cite{Ibanez:2012zk}, 
\begin{equation}
I_\alpha(q) = \int_k {\Gamma}_{\alpha\mu\nu}\Delta^{\mu\sigma}(k+q)\Delta^{\nu\rho}(k)B_{\rho\sigma} + \cdots ,
\label{Iint}
\end{equation}
where the ellipses indicate the graphs omitted in Fig.~\ref{uexpansion} (see~\cite{Aguilar:2011xe} for the complete version); their inclusion does not modify the basic argument, it simply makes it lengthier. In the  equations above we have introduced the dimensional regularization measure \mbox{$\int_k=\mu^\epsilon\int\!\diff{d} k/(2\pi)^d$} where $d=4-\epsilon$ is the space-time dimension and $\mu$ the 't Hooft mass.

We now renormalize the effective vertex $B$ by 
introducing the renormalization constant $Z_B$, 
\be
B_{\chic R}^{\rho\sigma} = Z_{\chic B}^{-1} B_0^{\rho\sigma},
\label{IB}
\ee
and combine \1eq{Uvert} and \1eq{Iint}. Since    
${\cal U}$ forms part of the three-gluon vertex (of the type $Q^3$), 
it renormalizes as   ${\cal U}^{\mu\alpha\beta}_{\chic R} = Z_3 {\cal U}^{\mu\alpha\beta}_0$, and with the help of \2eqs{renprop}{renconst2}, one 
concludes that (note that the dependence of $U$ on $B$ is effectively quadratic) 
\be
Z_{\chic B} = Z^{-1}_{\chic A}.
\label{ZBZA}
\ee

With these ingredients at hand, 
we now turn to the basic formula relating the gluon mass with the transition amplitude~\cite{Aguilar:2011xe,Ibanez:2012zk}, namely
\be
m(q^2) = g I(q^2).
\label{mgi}
\ee
Let us consider \1eq{mgi} written in terms of unrenormalized quantities, and 
substitute in its rhs the corresponding renormalized ones, by introducing 
the appropriate renormalization constants. Suppressing Lorentz indices and using  
$Z_g = Z_3  Z_{\chic A}^{-3/2}$ [see \1eq{STIrel}], one finds  
\be
m_0(q^2) = Z_{\chic A}^{-1/2} g_{\chic R}
\int_k {\Gamma}_{\chic R }\Delta_{\chic R}(k+q)\Delta_{\chic R}(k)B_{\chic R},
\label{mgir}
\ee
which clearly requires the renormalization dictated by \1eq{glmassren} in order to 
be converted into the manifestly renormalized form 
\be
m_{\chic R}(q^2) = g_{\chic R} I_{\chic R}(q^2).
\label{mgirf}
\ee

\subsection{The basic RGI quantities}

Let us finally consider certain RGI combinations of Green's functions, that will be useful in the ensuing analysis.
We recall that, by definition, a RGI combination maintains exactly the same form 
when written in terms of unrenormalized or renormalized quantities. 

To begin with, as is well-known, and easy to verify directly using~\1eq{renBFM},  the combination~\cite{Aguilar:2009nf,Aguilar:2010gm} 
\be
{d}(q^2) = g^2 {\widehat\Delta}(q^2)
=  \frac{g^2\, \Delta(q^2)}{[1+G(q^2)]^2},
\label{BRGI}
\ee
is an RGI quantity (note that in the second equality the BQI of \1eq{BQI} was employed).
It is then natural to {\it define} a RGI gluon mass, to be denoted by
$\overline{m}(q^2)$~\cite{Binosi:2012sj}, as 
\be
{\overline m}^2(q^2) =  g^{-2} {\widehat m}^2(q^2) 
= g^{-2} [1+G(q^2)]^2 m^2(q^2).
\label{mRGI}
\ee
We emphasize that the ${\overline m}^2(q^2)$ defined above is a convenient quantity to introduce, because, 
as will become apparent in the rest of this work, it helps us quantify the 
faithfulness of certain approximations with respect to the RG. 
Note, however, that no special physical meaning 
is ascribed to $\overline{m}(q^2)$ at this stage; in particular, despite its RGI nature we explicitly refrain from 
promoting it to a physical observable, for the simple reason that, at least within our present 
understanding, it is a quantity that depends on the gauge-fixing parameter. Specifically, all  
recent work related to the gluon mass equation has been performed in the Landau gauge, 
mainly because the corresponding lattice simulations have been carried out in this privileged gauge. 
In fact, the question whether the gluon propagator 
continues to saturate in the 
infrared when computed away from the Landau gauge
is practically unexplored, both on the lattice as well as within the SDE framework. 

We finally point out  that the definition of the RGI gluon mass introduced here differs from the alternative 
proposed in~\cite{Binosi:2013rba}, namely ${\overline m}^2(q^2)= m^2(q^2) J^{-1}(q^2)$.  
The problem with this latter definition is that, while formally RGI, gives rise 
to an ill-defined expression,  
due to the singular behavior of the quantity $J(q^2)$. Specifically, 
the contribution of the  massless ghost loop forces $J(q^2)$
to reverse its sign and finally diverge logarithmically in the deep infrared~\cite{Aguilar:2013vaa}; of course, the 
combination $q^2J(q^2)$, appearing in the definition of $\Delta^{-1}(q^2)$ [see \1eq{massive}] is perfectly finite.

Let us next introduce three additional RGI quantities, to be generically denoted by ${\cal R}_i$, formed out of special combinations of propagators, vertices, and the gauge coupling constant. 
In particular, we define 
\bea 
{\cal R}_1^{\mu\alpha\beta}(q,r,p) &=&  g \Delta^{1/2}(q)\Delta^{1/2}(r)\Delta^{1/2}(p) \Gamma^{\mu\alpha\beta}(q,r,p),
\nonumber\\
{\cal R}_2^{\mu}(q,r,p) &=&  g \Delta^{1/2}(q)^{1/2}(r) D^{1/2}(p) \Gamma^{\mu}(q,r,p),
\nonumber\\
{\cal R}_3^{\mu\alpha\beta\nu} (q,r,p,s)&=&  g^2  \Delta^{1/2}(q)\Delta^{1/2}(r)\Delta^{1/2}(p)
\Delta^{1/2}(s) \Gamma^{\mu\alpha\beta\nu}(q,r,p,s).
\label{R1R2R3}
\eea
The RGI nature of the above quantities may be verified directly, by employing the relations listed in \1eq{STIrel}.  

\section{\label{RGprop} RG properties of the full gluon mass equation}


\begin{figure}[!t]
\includegraphics[scale=0.9]{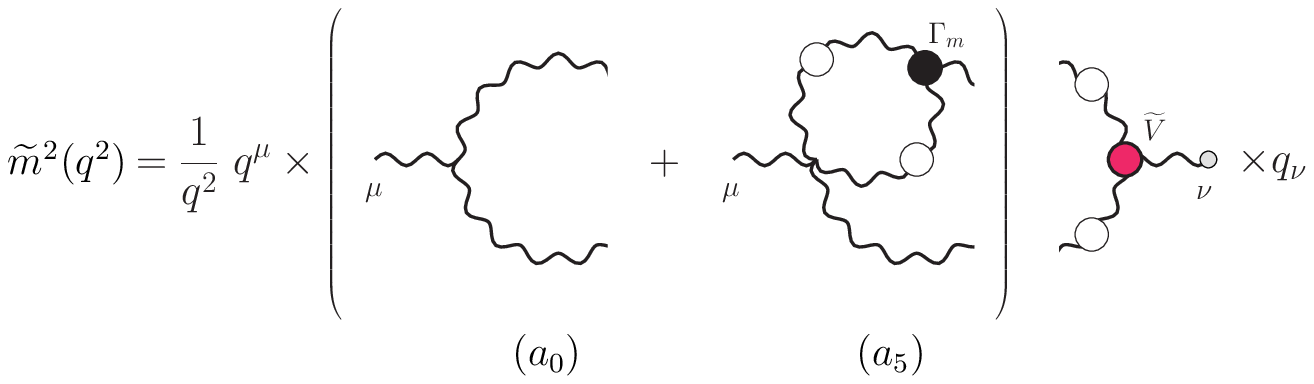}  
\caption{\label{massfig} Diagrammatic representation of the gluon mass equation.}
\end{figure}


In this section we study the RG structure of the integral equation that 
controls the momentum evolution of the gluon mass. The main result of this analysis
is that the {\it complete} kernel of this equation acquires a form that allows 
both of its sides to be written in terms of the RGI quantities introduced in the previous section.

\begin{figure}[!t]
\mbox{}\hspace{-0.7cm}\includegraphics[scale=0.67]{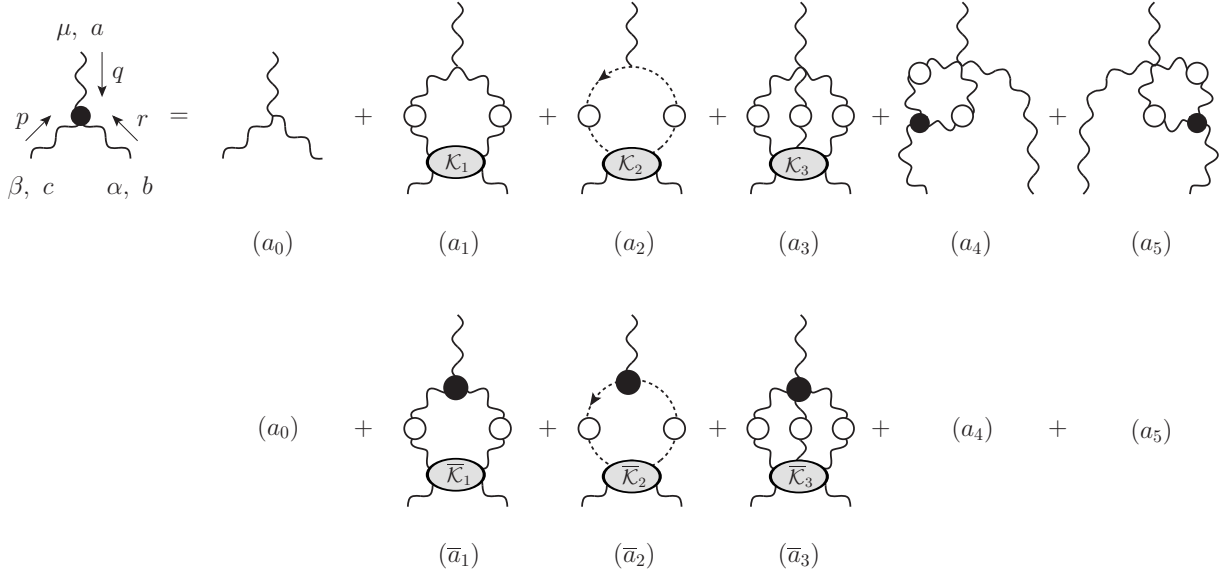}  
\caption{\label{vertex} The SDE for the three gluon vertex, in the conventional (first row) and Bethe Salpeter version (second row).
Note that the Bose symmetry of $\Gamma_{\mu\alpha\beta}^{abc}(q,r,p)$ implies that 
\mbox{$(a_4)_{\mu\alpha\beta}^{abc}(q,r,p)=(a_5)_{\mu\beta\alpha}^{acb}(q,p,r)$}; when the color has been factored out, 
as in \1eq{masseq2}, we have instead \mbox{$(a_4)_{\mu\alpha\beta}(q,r,p)=-(a_5)_{\mu\beta\alpha}(q,p,r)$}.}
\end{figure}

As has been demonstrated in~\cite{Binosi:2012sj}, the complete gluon mass equation is given by (see  Fig.~\ref{massfig}) 
\be 
m^2(q^2) =\frac{1}{2}\, \frac{i\gA}{1+G(q^2)} \frac{q_\mu q_\nu}{q^2} 
\int_k\! [(a_0) +2 (a_5)]^{\mu\alpha\beta} \Delta_{\alpha\rho}(k)\Delta_{\beta\sigma}(k+q) {\NP}^{\nu\rho\sigma}(q,k,-k-q),
\label{masseq1}
\ee
where $C_A$  is the Casimir eigenvalue in the adjoint representation [$C_A=N$ for $SU(N)$], ${\NP}$ is the pole 
vertex introduced in the previous section, $(a_0)$ is simply the 
tree-level three-gluon vertex, 
\be
(a_0)_{\mu\alpha\beta} = \gtree_{\mu\alpha\beta}(q,k,-k-q),
\label{verttree}
\ee
with
\be
\Gamma_{\mu\alpha\beta}^{(0)}(q,r,p) = 
(q-r)_{\beta}g_{\alpha\mu} + (r-p)_{\mu}g_{\alpha\beta} + (p-q)_{\alpha} g_{\mu\beta} ,
\ee
and $(a_5)$ denotes the vertex subgraph nested in 
the ``two-loop'' self-energy graph (see also \fig{vertex}). 

Using the fact that ${\NP}$ satisfies the WI of~\1eq{winp} \ie
\be
q^{\nu} \NP_{\nu\rho\sigma}(q,k,-k-q)= m^2(k)P_{\rho\sigma}(k) - m^2(k+q)P_{\rho\sigma}(k+q), 
\label{winp1}
\ee
and after appropriate shifts of the integration variable, we arrive at~\cite{Binosi:2012sj}
\be 
m^2(q^2) = \frac{i\gA}{1+G(q^2)} \frac{q_\mu}{q^2} 
\int_k\! [(a_0) +(a_4)+ (a_5)]^{\mu\alpha\beta} \Delta_{\alpha\rho}(k)\Delta_{\beta}^{\rho}(k+q) m^2(k^2),
\label{masseq2}
\ee
where $(a_4)_{\mu\alpha\beta}(q,r,p)=-(a_5)_{\mu\beta\alpha}(q,p,r)$ (see also the first row of \fig{vertex}). 


Let us now turn to the SDE satisfied by the conventional ($Q^3$) three gluon vertex $\Gamma_{\mu\alpha\beta}^{abc}(q,r,p)$, 
shown diagrammatically in \fig{vertex}, and derive 
a relation necessary for the treatment of \1eq{masseq2}. 
On the first line of \fig{vertex},  
the vertex SDE is expressed in terms of the standard multiparticle kernels, ${\cal K}_i$,  
while on the second the Bethe-Salpeter version of the same equation is presented. 
Note that in this latter version the vertices with the external momentum $q$ are fully 
dressed; consequently,  
the corresponding Bethe-Salpeter kernels,  $\overline{\cal K}_i$ differ from the ${\cal K}_i$, 
since certain diagrams, allotted to dress the vertices, must be excluded from them, in order to 
avoid overcounting ($\overline{\cal K}_i$ and ${\cal K}_i$ are related through a non-linear integral equation 
(see, \eg~\cite{Roberts:1994dr} and~\cite{Bjorken:1979dk}). 

If we express the various diagrams $(a_i)$ 
in terms of renormalized quantities, denoting by $(a_i^{\chic R})$ the resulting expressions, 
it is relatively straightforward to demonstrate that
\begin{align}
({\overline a}_1) &= Z_3 ^{-1} ({\overline a}_1^{\chic R});&
({\overline a}_2)   &= Z_3 ^{-1} ({\overline a}_2^{\chic R});&
({\overline a}_3) &= Z_3 ^{-1} ({\overline a}_3^{\chic R});
\nonumber \\
(a_4) &= Z_g^2   Z_{\chic A}^{2} Z_3 ^{-1} (a_4^{\chic R});&
(a_5) &= Z_g^2   Z_{\chic A}^{2} Z_3 ^{-1} (a_5^{\chic R}).
\end{align}

Thus, for the original, unrenormalized vertex SDE we have (suppressing all indices) 
\bea   
\Gamma &=& (a_0) + (a_4) + (a_5) + ({\overline a}_1) + ({\overline a}_2) + ({\overline a}_3) 
\nonumber\\
&=& (a_0) + Z_g^2   Z_{\chic A}^{2} Z_3 ^{-1}[(a_4^{\chic R})+(a_5^{\chic R})] + 
Z_3 ^{-1} [({\overline a}_1^{\chic R}) + ({\overline a}_2^{\chic R}) + ({\overline a}_3^{\chic R})],
\label{twoside}
\eea
and so, after introducing the renormalized vertex $\Gamma_{\chic R} = Z_3 \Gamma$ [see \1eq{renconst2}], we arrive at
\be
Z_3( a_0) + Z_g^2   Z_{\chic A}^{2} \left[(a_4^{\chic R})+(a_5^{\chic R})\right] = 
\Gamma_{\chic R} - [({\overline a}_1^{\chic R}) + ({\overline a}_2^{\chic R}) + ({\overline a}_3^{\chic R})].
\label{airen}
\ee

Returning to \1eq{masseq2}, and rewriting it in terms of renormalized quantities, we have 
\be 
m^2_{\chic R}(q^2) = \frac{i C_A g^2_{\chic R}} {1+G_{\chic R}(q^2)} \frac{q_\mu}{q^2} 
\int_k\! \Big\{Z_3( a_0) + Z_g^2   Z_{\chic A}^{2} [(a_4^{\chic R})+(a_5^{\chic R})]\Big\}^{\mu\alpha\beta} 
\Delta^{\chic R}_{\alpha\rho}(k)\Delta^{\chic R \,\rho}_{\beta}(k+q)\, m^2_{\chic R}(k^2),
\label{mren1}
\ee
which, in view of \1eq{airen}, may be written exclusively in terms of renormalized quantities 
(\ie with no reference to the cutoff-dependent $Z_i$), as 
\be 
m^2_{\chic R}(q^2) = \frac{i C_A g^2_{\chic R}} {1+G_{\chic R}(q^2)} \frac{q_\mu}{q^2} 
\int_k \,{\cal G}^{\mu\alpha\beta} 
\Delta^{\chic R}_{\alpha\rho}(k)\Delta^{\chic R \,\rho}_{\beta}(k+q)\, m^2_{\chic R}(k^2),
\label{mren2}
\ee
where
\be
{\cal G}^{\mu\alpha\beta}
\equiv  
\bigg[\Gamma_{\chic R} -\sum_{i=1}^{3} ({\overline a}_i^{\chic R})\bigg]^{\mu\alpha\beta} ,
\ee
namely the rhs of \1eq{airen}.

We next study the properties of \1eq{mren2} under RG transformations. To that end, it is convenient  
to recast both sides of this equation in terms of appropriately chosen RGI quantities. 
Clearly, by virtue of \1eq{mRGI}, 
a simple multiplication by $g^{-2}[1+G(q^2)]^2$
converts the lhs of \1eq{mren2} into the RGI mass 
${\mrgi}^2 (q^2)$ introduced in \1eq{mRGI}. On the other hand, 
the demonstration that, after the aforementioned multiplication,  the rhs  is also RGI, is slightly more involved.


\begin{figure}[!t]
\includegraphics[scale=0.625]{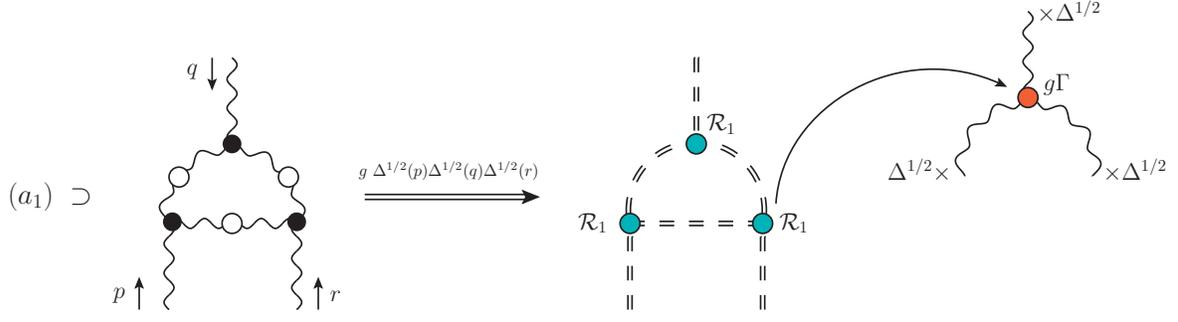}  
\caption{\label{RGIness} Schematic representation of the conversion of a typical diagram into its RGI equivalent.}
\end{figure}

To prove this statement, 
we will employ the three RGI quantities, ${\cal R}_i$, introduced in \1eq{R1R2R3}.
In particular, it is relatively straightforward to establish that when 
the terms $({\overline a}_i^{\chic R})$ 
are multiplied by the factor $g_{\chic R} \Delta_{\chic R}^{1/2}(q) \Delta_{\chic R}^{1/2}(r) \Delta_{\chic R}^{1/2}(p)$
they become functions of the ${\cal R}_i$; so, we have (see Fig.\ref{RGIness})
\be
g_{\chic R} \Delta_{\chic R}^{1/2}(q) \Delta_{\chic R}^{1/2}(r) \Delta_{\chic R}^{1/2}(p)({\overline a}_i^{\chic R}) 
={\cal F}_i({\cal R}_1,{\cal R}_2,{\cal R}_3).
\label{aux1}
\ee
As a consequence, 
\be
g_{\chic R} \Delta_{\chic R}^{1/2}(q) \Delta_{\chic R}^{1/2}(r) \Delta_{\chic R}^{1/2}(p)\, {\cal G}(q,r,p) 
= {\cal R}_1 - \sum_{i=1}^{3}{\cal F}_i \equiv {\cal R}.
\label{GRGI}
\ee
Note finally that the ratio $f(p_1)/f(p_2)$ 
of any two-point function $f(p)$ is also a RGI quantity. 

Armed with these results, 
we may now re-express \1eq{mren2} in terms of manifestly RGI quantities. 
Specifically, after the aforementioned multiplication by $g^{-2}[1+G(q^2)]^2$, 
and some appropriate manipulations, 
we arrive at [with $p = -(k+q)$] 
\be 
\overline{m}^2(q^2) = \frac{i C_A \,q_\mu}{q^2  {d}^{1/2}(q^2)} 
\int_k{\cal R}^{\mu\alpha\beta}  P_{\alpha\rho}(k) P^{\rho}_{\beta}(p) {d}^{1/2}(k^2){d}^{1/2}(p^2)
\left\{\frac{1+G(p^2)}{1+G(k^2)}\right\}  \overline{m}^2(k^2),
\label{mrgi}
\ee
which is a manifestly RGI integral equation.

\section{\label{tenstr} General structure of the three-gluon kernel}

In the previous section we have demonstrated that the mass equation, as captured in \1eq{mren2}, 
has built in it the exact RG properties that one expects on general theoretical grounds. 
Evidently, in order to proceed further, and deduce from \1eq{mren2} the  
momentum dependence of the gluon mass, 
further information on ${\cal G}^{\mu\alpha\beta}$, or directly on its 
divergence, $q_{\mu}{\cal G}^{\mu\alpha\beta}$, is needed.  

It is clear that the diagrammatic decomposition of ${\cal G}^{\mu\alpha\beta}$ involves 
the three Bethe-Salpeter kernels  $\overline{\cal K}_1$, $\overline{\cal K}_2$, and 
$\overline{\cal K}_3$, 
whose complicated skeleton expansion renders their full determination  
impossible. It is therefore necessary, for practical purposes, to introduce approximations or Ans\"atze
for the quantity ${\cal G}^{\mu\alpha\beta}$, which ought to 
encode, as well as possible, some of its  
salient field-theoretic properties.

To that end, it is essential to consider the 
tensorial decomposition of ${\cal G}^{\mu\alpha\beta}$, and exploit its 
Bose-symmetric nature, together with the fact that, 
when inserted into \1eq{mren2}, it is contracted by two transverse projectors.   
Specifically, in a straightforward basis
composed by the momenta $r$ and $p$, one has\footnote{Alternatively, one may use the 
standard Ball and Chiu decomposition~\cite{Ball:1980ay}, arriving at exactly the same conclusions.}  

\be
{\cal G}^{\mu\alpha\beta}(q,r,p) = \sum_{i=1}^{14} C_i(q,r,p) b_i^{\mu\alpha\beta},
\ee
with 
\begin{align}
b_1^{\mu\alpha\beta} & = r^{\mu} g^{\alpha\beta};&
b_2^{\mu\alpha\beta} & = p^{\mu} g^{\alpha\beta} ;&
b_3^{\mu\alpha\beta}  &=  p^{\alpha} g^{\mu\beta};
\nonumber\\
b_4^{\mu\alpha\beta} &= r^{\beta} g^{\mu\alpha};&
b_{5}^{\mu\alpha\beta} &= p^{\mu} p^{\alpha} r^{\beta};&
b_{6}^{\mu\alpha\beta} &= r^{\mu} p^{\alpha} r^{\beta},
\label{bs1}
\end{align}
and 
\begin{align}
b_7^{\mu\alpha\beta} &= p^{\beta} g^{\mu\alpha};&
b_8^{\mu\alpha\beta} &= r^{\alpha} g^{\mu\beta};&
b_9^{\mu\alpha\beta} &= r^{\mu} r^{\alpha} r^{\beta};&
b_{10}^{\mu\alpha\beta} &= p^{\mu} p^{\alpha} p^{\beta};
\nonumber\\
b_{11}^{\mu\alpha\beta} &= p^{\mu} r^{\alpha} p^{\beta};&
b_{12}^{\mu\alpha\beta} &= p^{\mu} r^{\alpha} r^{\beta};&
b_{13}^{\mu\alpha\beta} &=  r^{\mu} p^{\alpha} p^{\beta};&
b_{14}^{\mu\alpha\beta} &= r^{\mu} p^{\alpha} p^{\beta}.
\label{bs2}
\end{align}
The form factors $C_i(q,r,p)$ are in general related among each other by conditions
imposed by Bose-symmetry. Particularly important to our purposes are the relations 
\be
C_2(q,r,p) = -C_1(q,p,r);\qquad  C_4(q,r,p) = - C_3(q,p,r).
\label{boserel}
\ee
The tree-level values of the form factors $C_i$ are determined by setting ${\cal G} = \gtree$; 
as one can check by substituting $q= -(r+p)$ into \1eq{verttree}, and using the above basis to express the result, on has
$C_1^{(0)}=1$, $C_2^{(0)}=-1$, $C_3^{(0)}=2$, $C_4^{(0)}=-2$, $C_7^{(0)}=-1$, $C_8^{(0)}=1$, with all the remaining $C$s vanishing.

Now, when contracted with $P_{\alpha\rho}(r) P^{\rho}_{\beta}(p)$, 
the second set of tensors, $(b_7)-(b_{14})$, vanishes identically, while for the first set, one can effectively use the replacements 
\begin{align}
b_3^{\mu\alpha\beta}  &\to - q^{\alpha} g^{\mu\beta};&
b_4^{\mu\alpha\beta} &\to -q^{\beta} g^{\mu\alpha};&
b_{5}^{\mu\alpha\beta} &\to p^{\mu} q^{\alpha} q^{\beta};&
b_{6}^{\mu\alpha\beta} &\to  r^{\mu} q^{\alpha} q^{\beta}.
\end{align}

It is then relatively straightforward to establish that 
\be
q_{\mu}{\cal G}^{\mu\alpha\beta} P_{\alpha\rho}(r) P^{\rho}_{\beta}(p)= 
\left\{[(p^2-r^2) S + q^2 A ]g^{\alpha\beta} + B q^{\alpha}q^{\beta} \right\}
P_{\alpha\rho}(r) P^{\rho}_{\beta}(p).
\label{divG}
\ee
where 
\bea
S(q,r,p) &=& \frac{1}{2} [C_1(q,r,p) + C_1(q,p,r)],
\nonumber\\
A(q,r,p) &=& \frac{1}{2} [C_1(q,p,r) - C_1(q,r,p)],
\nonumber\\
B(q,r,p) &=& (q\cdot p)\, C_5(q,r,p) + (q\cdot r)\, C_6(q,r,p) + [C_3(q,p,r)-C_3(q,r,p)].
\label{SAB}
\eea
The terms $A$ and $B$ emerge when writing the total contribution from $b_1$ and  $b_2$ 
as the sum of a symmetric and an antisymmetric piece under $r\leftrightarrow p$, namely
\be
[C_1(q,r,p) r^{\mu} + C_2(q,r,p) p^{\mu}]g^{\alpha\beta} = [S(q,r,p) (r-p)^{\mu} + A(q,r,p) q^{\mu}]g^{\alpha\beta}.
\label{sa}
\ee
In addition, note that we have used \1eq{boserel} to eliminate $C_2$ and $C_4$ in favor of $C_1$ and $C_3$, respectively.

Let us now comment on the way that the terms of \1eq{SAB} contribute to the mass equation in the limit $q\to 0$. 
It is easy to verify that the terms associated with $A(q,r,p)$ and $B(q,r,p)$ 
are subleading in this limit. 
Indeed, first of all, the $q^2 $ that multiplies the $A(q,r,p)$ and the  
$q^{\alpha}q^{\beta}$ that multiplies $B(q,r,p)$ compensate the $(1/q^2)$ in front 
of the mass equation.
Then, since $A(q,r,p)$ is antisymmetric under $r\leftrightarrow p$, we have that $A(0,-p,p) = 0$, and therefore, 
when $q\to 0$, \mbox{$A(q,r,p) \to {\cal O}(q)$}.
Similarly, 
the terms in $B$ proportional to $C_5$ and $C_6$ are multiplied by an additional 
power of $q$, and are manifestly subleading, while the remaining term 
is antisymmetric under $r\leftrightarrow p$, and therefore this too is of order ${\cal O}(q)$.
Thus, the only term that contributes to the mass equation in the IR limit is the 
one associated with $S$.
Note finally that out of the three terms defined in \1eq{SAB}, only $S$ has a tree level value, 
namely $S^{(0)}=1$.

After these considerations, we can write down the final form taken by the mass equation. 
Setting $r=k$, $p=-(k+q)$, and passing to Euclidean space following standard rules~\cite{Binosi:2012sj}, 
(and suppressing the index ``E'' throughout), we have 
\be
m^2(q^2) = -\frac{g^2 C_A}{1+G(q^2)}\frac{1}{q^2}\int_k m^2(k^2) \Delta_{\alpha\rho}(k)\Delta^{\rho}_{\beta}(k+q) 
{\cal K}^{\alpha\beta}(q,k,-k-q),
\label{masseqnew}
\ee
where, according to the above discussion, the total kernel ${\cal K}^{\alpha\beta }$ may be naturally 
decomposed into a contribution that is leading in the IR, to be denoted by 
${\cal K}^{\alpha\beta}_{\rm{L}}$, and one that is subleading, to be 
denoted by ${\cal K}^{\alpha\beta}_{\rm{SL}}$, namely 
\be
{\cal K}^{\alpha\beta} = {\cal K}^{\alpha\beta}_{\rm{L}} + {\cal K}^{\alpha\beta}_{\rm{SL}},
\label{sumK}
\ee
with 
\bea
{\cal K}^{\alpha\beta}_{\rm{L}} &=& [(k+q)^2 - k^2] S g^{\alpha\beta}, 
\nonumber\\
{\cal K}^{\alpha\beta}_{\rm{SL}} &=& q^2 A g^{\alpha\beta} + B q^{\alpha}q^{\beta},
\label{theKs}
\eea
where the common argument $(q,k,-q-k)$ in all above quantities has been suppressed.

\section{\label{RGorig} RG properties of the original mass equation}

Let us now compare \1eq{masseqnew} with the one derived originally in~\cite{Binosi:2012sj}.
There, the mass equation considered 
had the form of \1eq{mren1}; in other words, one dealt directly with  
the term  $[Z_3( a_0) + Z_g^2   Z_{\chic A}^{2} [(a_4^{\chic R})+(a_5^{\chic R})]$, without 
passing to the rhs of \1eq{airen}. 
The way to handle the renormalization constants 
was to set them directly equal to unity, 
and {\it assume} that the remaining terms had been rendered UV finite.
This procedure finally amounts to the effective replacement 
\be
q_{\mu}\bigg\{Z_3( a_0) + Z_g^2   Z_{\chic A}^{2} [(a_4^{\chic R})+(a_5^{\chic R})]\bigg \}^{\mu\alpha\beta} 
\to  {\cal K}^{\alpha\beta}(k,q),
\ee
with 
\bea
{\cal K}^{\alpha\beta}(k,q) &=& [(k+q)^2 - k^2] \left\{ 1 - [\Y_{\chic R}(k+q) + \Y_{\chic R}(k)]\right\}g^{\alpha\beta}
\nonumber\\
&+& [\Y_{\chic R}(k+q)-\Y_{\chic R}(k)](q^2 g^{\alpha\beta}-2q^{\alpha} q^{\beta}),
\label{massK}
\eea
where 
\be
\Y(k^2)=\frac{g^2 C_A}{4 k^2} \,k_\alpha\! \int_\ell\!\Delta^{\alpha\rho}(\ell)
\Delta^{\beta\sigma}(\ell+k)\Gamma_{\sigma\rho\beta}(-\ell-k,\ell,k). 
\label{theY}
\ee
The renormalized version of $\Y$ is simply   
\be
Y_{\!\s{R}}(k) = Y(k) - Y(\mu), 
\ee
namely the form corresponding to the momentum-subtraction (MOM) scheme.
 
A direct comparison of \1eq{massK} with the generic form given in \1eq{theKs} establishes that, 
in this case,  
\begin{align}
S &= 1-[\Y_{\!\s{R}}(k+q) + \Y_{\!\s{R}}(k)];&
A &= \Y_{\chic R}(k+q)-\Y_{\chic R}(k);& 
B = -2A.
\label{SABapp}
\end{align}
Note that $S$ is symmetric under the interchange 
$k \leftrightarrow (k+q)$, as expected from its general property  given in \1eq{SAB}; similarly, the $A$ of \1eq{SABapp}
is antisymmetric under the same interchange, exactly as the $A$ of \1eq{SAB}. 
Finally, $S^{(0)}=1$, as it should.  

In~\cite{Binosi:2012sj} an approximate form for $\Y(k)$ was 
obtained by substituting tree-level expressions  for all quantities 
appearing inside the integral in \1eq{theY}.
The result is given by 
\be
Y_{\!\s{R}}(k^2)= - \frac{15}{16} \,t(k) ,
\label{Yappr}
\ee
where 
\be
t(k)\equiv \left(\frac {\alpha_s C_A}{4\pi}\right) \log\left(\frac{k^2}{\mu^2}\right),
\label{tdef}
\ee
and $\alpha_s = g^2/4 \pi$ is the value of the Yang-Mills charge 
at the subtraction point $\mu$ chosen.

In the analysis of the gluon mass equation 
presented in \cite{Binosi:2012sj}, the rhs of \1eq{Yappr} was multiplied by a constant $C$, with $C>1$. 
As has been explained in detail there, 
the main reason for this is the need to counteract the (destabilizing) effect of the  
negative sign in front of the integral on the rhs of \1eq{masseqnew}, 
and obtain positive-definite solutions for the gluon mass, at least within a reasonable range 
of physical momenta. 
In particular, for $\alpha_s=0.22$, which is the ``canonical'' MOM value for \mbox{$\mu=4.3$ GeV}, 
and $C=9.2$, the function  
$m^2(q^2)$ is positive in the range of momenta  between 0 to \mbox{$5.5$ GeV} ; past that point 
it turns negative (but its magnitude is extremely small, around $10^{-5}\, \mbox{GeV}^2$, as shown in the inset
of Fig.~\ref{fig:comp_mass})~\cite{Holdom:2013uya}. As we will see in the next section, 
this unwanted feature may be eventually rectified, 
by modifying appropriately the form of $S$.  

\subsection{Quantifying the kernel quality: The basic procedure}

In order to quantitatively determine 
to what extent a given approximation for $S$ respects the RG properties 
of the full mass equation,  
it is necessary to establish a reference situation, and then compute possible deviations from it. 
To that end, we will 
employ a general procedure that consists of the following main steps.

\n{\it i}
We consider the RGI quantity 
\be
d(q^2) = g^2\, F^2(q^2) \Delta(q^2),
\label{BRGIF}
\ee
namely that of \1eq{BRGI} with the approximation \1eq{GFapp} implemented, 
and compute its shape using $F(q^2)$ and $\Delta(q^2)$ from the lattice, for different 
values of the renormalization point $\mu$. To that end, we  
use the standard formulas~\cite{Aguilar:2009nf} 
\be
\Delta(q^2,\mu^2)=\frac{\Delta(q^2,\nu^2)}{\mu^2\Delta(\mu^2,\nu^2)}, 
\qquad F(q^2,\mu^2)=\frac{F(q^2,\nu^2)}{F(\mu^2,\nu^2)},  
\label{ren_gl}
\ee
which allow one to  
connect a set of lattice data renormalized at $\mu$ with the corresponding set renormalized at $\nu$.
It is clear that, since these changes amount 
to the multiplication of the 
product $F^2(q^2) \Delta(q^2)$ by an overall constant, 
we can adjust the value of $g^2$ (or $\alpha_s$) for each $\mu$, 
such that the curves of $d(q^2)$ so produced lie exactly on top of each other. 
Thus, this procedure fixes the values of $\alpha_s (\mu)$, such that, the 
(formally RGI) $d(q^2)$ is indeed RGI. As we will see, the resulting values for $\alpha_s (\mu)$
are rather compatible with those predicted by standard MOM calculations. 

\n{\it ii}
We next solve the 
gluon mass equation for the same set of $\mu$'s used in the previous step. Specifically, 
for the ingredients entering in the rhs of \1eq{masseqnew}, such as $g^2$, $F$, and $\Delta$, 
we use the corresponding quantities found in \n{\it i}, for any given $\mu$; note that 
$Y_{\!\s{R}}$ is also $\mu$-dependent, and is accordingly modified. 
This procedure furnishes a set of $\mu$-dependent solutions, $m^2(q^2,\mu^2)$;  
note that the value of the constant $C$ that multiplies $Y_{\!\s{R}}$ 
also varies (rather mildly) with $\mu$.

\n{\it iii}
The various masses,  $m^2(q^2,\mu^2)$, found in \n{\it ii} are now used 
to construct the  RGI mass defined in \1eq{mRGI} [using again \1eq{GFapp}], namely 
\be
{\overline m}^2(q^2) = \frac{m^2(q^2)}{g^{2} F^2(q^2)}.
\label{mRGIF}
\ee
Now, ideally speaking, when the various  $m^2(q^2,\mu^2)$ are inserted into \1eq{mRGIF}, 
together with the corresponding ($\mu$-dependent) $g^{2} F^2(q^2)$, one ought to 
obtain the same identical curve for each value of $\mu$.

In practice, of course, deviations between the various curves are expected, precisely 
because our knowledge of $S$ is imperfect. Therefore, a theoretically motivated way to discriminate 
between possible approximation for $S$ is to choose the one that 
produces the best coincidence (in the sense of minimizing the relative error) 
for the various ${\overline m}^2(q^2)$.

\subsection{Numerical analysis} 

Throughout the numerical study presented here, as well as in the next section, we will 
evaluate the relevant field-theoretic quantities at three different values of the renormalization point $\mu_i$; in particular,
we will use \mbox{\mbox{$\mu_1=4.3$ GeV}}, \mbox{\mbox{$\mu_2=3.0$ GeV}}, and \mbox{\mbox{$\mu_3=2.5$ GeV}}.
In the various plots, the curves of a quantity $A(q^2,\mu_i^2)$ produced for these three different values of $\mu_i$ 
 will be 
depicted as follows: $A(q^2,\mu_1^2)$ with squares or solid (black) curve; 
$A(q^2,\mu_2^2)$ with circles or dotted (red) curve; $A(q^2,\mu^2_3)$ with triangles or dashed (blue) curve.

\begin{figure}[!t]
\hspace{-1.5cm}\includegraphics[scale=1]{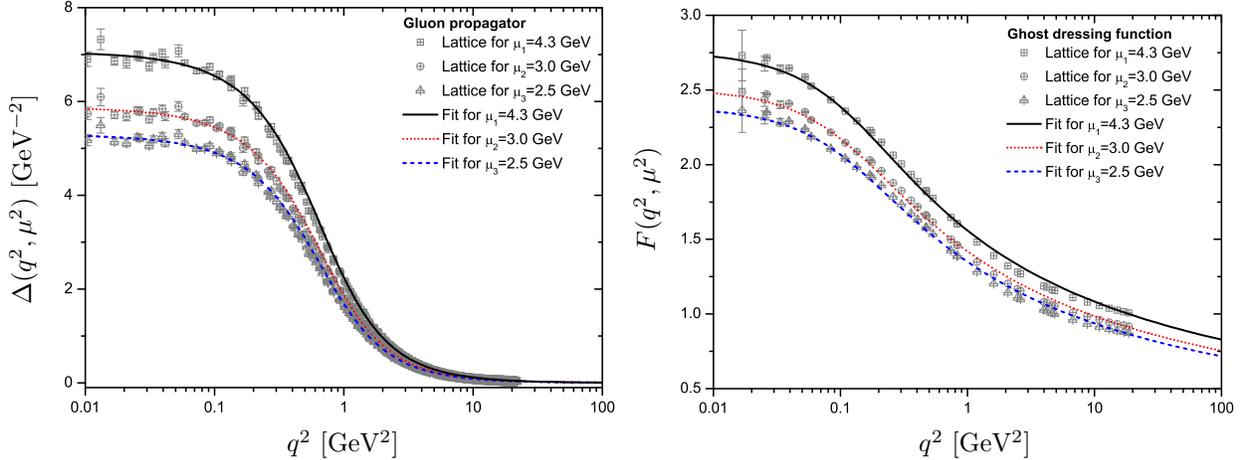}
\caption{\label{fig:prop} (color online) The quenched lattice data and the corresponding fits for the $SU(3)$ gluon propagator (left panel) and ghost dressing function (right panel) renormalized  at three different scales $\mu_i$. Lattice data are taken from~\cite{Bogolubsky:2009dc}.}  
\end{figure}

The first step in this analysis is to 
consider the lattice data for $\Delta(q^2)$  and $F(q^2)$ given in~\cite{Bogolubsky:2009dc}; 
these data are fitted using 
the functional forms reported in various recent articles~\cite{Aguilar:2010cn,Aguilar:2011ux,Aguilar:2012rz}. 
Then, repeated use of \1eq{ren_gl} allows us to 
generate the three curves for $\Delta(q^2)$  and $F(q^2)$ renormalized at $\mu_i$ (with $i=1,2,3$), 
which are shown in~\fig{fig:prop}.  
It is clear that, due to multiplicative renormalizability, expressed by~\1eq{ren_gl},  
each curve may be obtained from the other by a simple 
rescaling. Specifically, the curves $\Delta(q^2,\mu_2^2)$  and $\Delta(q^2,\mu_3^2)$ are 
obtained from $\Delta(q^2,\mu_1^2)$ through multiplication by the factors of $1.20$ and $1.33$, respectively;
in the case of $F(q^2)$, the corresponding rescaling factors are $1.09$ and $1.15$.


Next, we form the RGI combination $d(q^2)$ given  
in~\1eq{BRGIF}. Concretely, for each specific value of $\mu_i$, we combine the corresponding 
ingredients entering into the definition of $d(q^2)$. As mentioned before in
step \n{\it i},  the value of $g^2$ (or $\alpha_s$)  for each $\mu_i$
is fixed by requiring that the three curves of $d(q^2)$ so produced lie exactly on top of each other;
so, the corresponding $\alpha_s(\mu_i)$ must be rescaled by an amount that will exactly compensate the 
corresponding rescalings introduced to the product $\Delta(q^2,\mu_i^2) F^2(q^2,\mu_i^2)$.  
Specifically, starting with $\alpha_s(\mu_1^2)=0.220$, which is the value that best fits the 
lattice data in the recent SDE analysis presented in~\cite{Aguilar:2013xqa}, we obtain the values  
$\alpha_s(\mu_2^2)=0.320$ and $\alpha_s(\mu_3^2)=0.392$.

\begin{figure}[!t]
\hspace{-1.5cm}\includegraphics[scale=1]{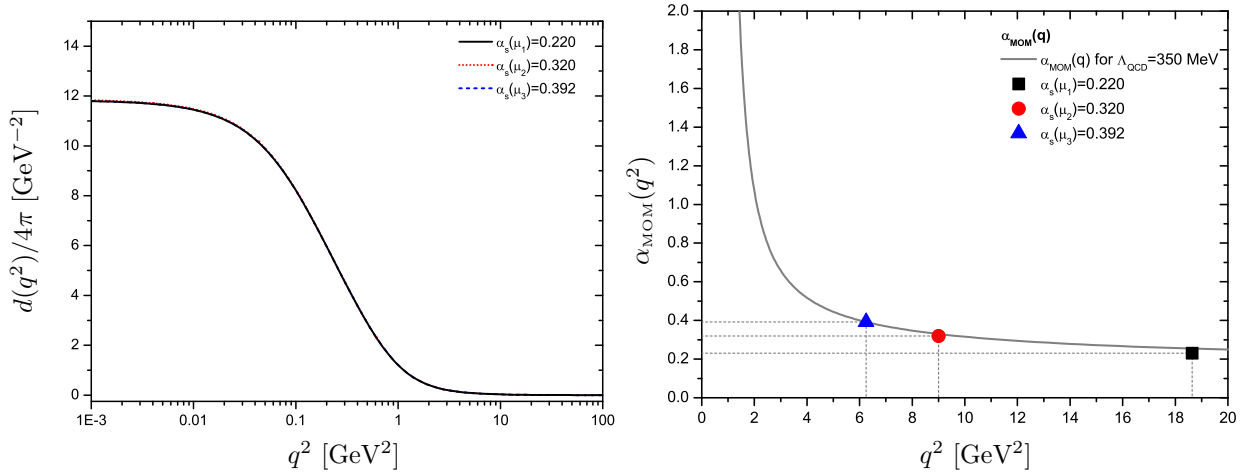}
\caption{\label{fig:RGI-alpha}(color online) Left panel: The RGI combination $d(q^2)/4\pi$  obtained using 
Eq.~(\ref{BRGI}) for the three renormalization points $\mu_i$ chosen. Right panel: The running coupling in the MOM scheme, $\alpha_\s{\rm MOM}(q)$~\cite{Boucaud:2008gn} for \mbox{$\Lambda_\s{\rm QCD}=350$ MeV} and $N_f=0$. Each point represent the values used for $\alpha_s(\mu_i)$ in our calculations.}
\end{figure}

On the left panel of~\fig{fig:RGI-alpha}  we plot the three curves for 
the dimensionful quantity  $d(q^2)/4\pi$. As expected, by construction, we can see 
that  the three curves  are indeed  on top of each other, thus making manifest that, 
for the particular set of values of  $\alpha_s(\mu_i)$  quoted above,  
$d(q^2)$ is $\mu$-independent.


It is important at this point 
to check whether the values for $\alpha_s(\mu_i)$ obtained from the above procedure are compatible
with the MOM expectations.  
This is done in the right panel of the same figure, where the gray continuous line 
represents the $\alpha_\s{\rm MOM} (q^2)$ 
obtained from the nonperturbative analysis of~\cite{Boucaud:2008gn}, for \mbox{$\Lambda_\s{\rm QCD}=350$ MeV} and $\Nf=0$; 
the aforementioned three values used for  $\alpha_s(\mu_i)$ are denoted by the corresponding symbols.  
As we can see, the values of $\alpha_s(\mu_i)$ that implement the $\mu$-independence of $d(q^2)$
are indeed in good agreement with the MOM predictions.

\begin{figure}[!t]
\hspace{-1.5cm}\includegraphics[scale=1]{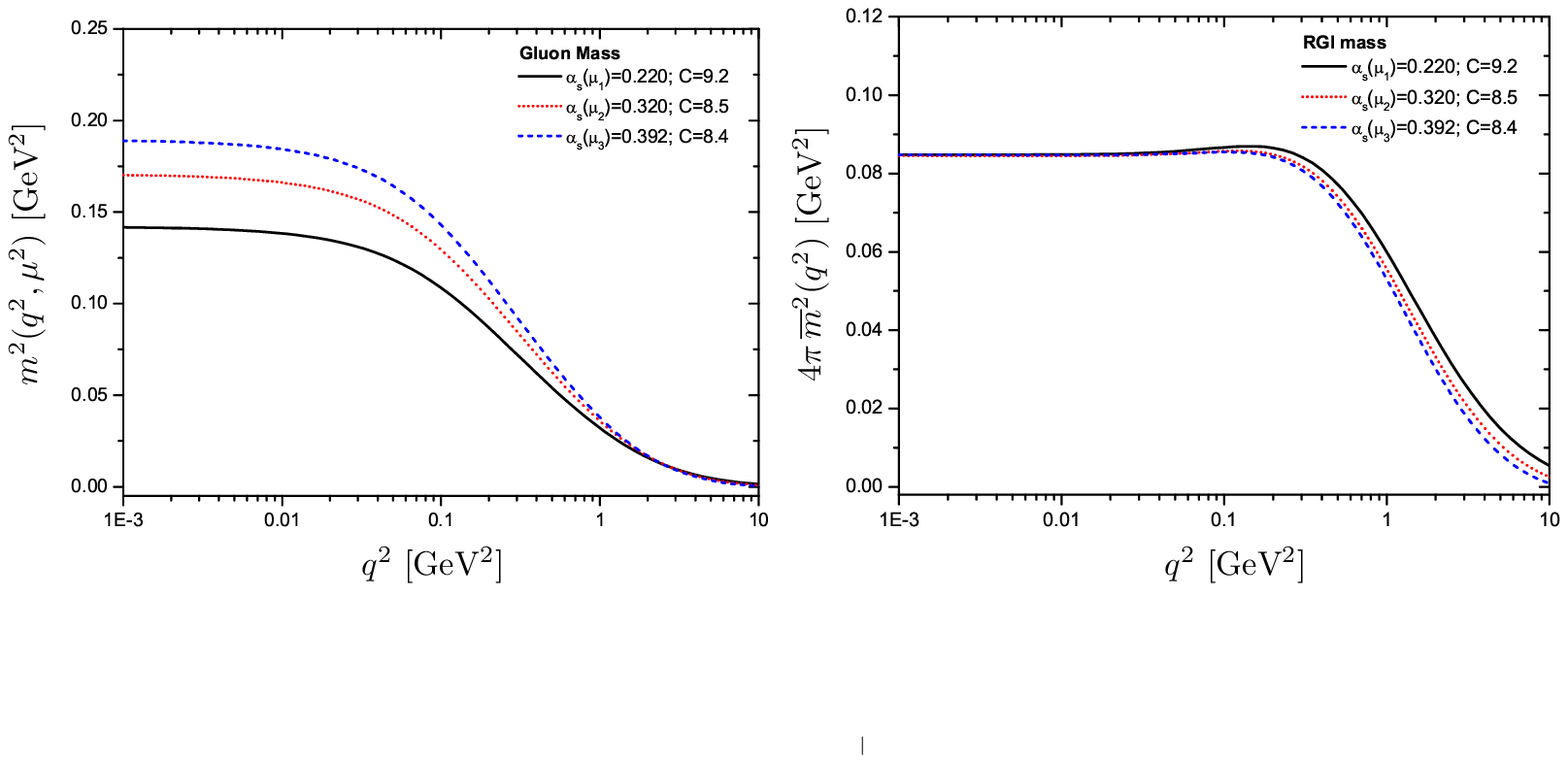}
\caption{\label{fig:m_or}(color online) Left panel: The numerical solution for the dynamical gluon mass, \mbox{$m^2(q^2,\mu^2)$},
for the three values of \mbox{$\mu_i$}. Right panel: The  corresponding RGI mass \mbox{$4\pi{\overline m}^2(q^2)$} obtained from~\1eq{mRGI} for  the three values of \mbox{$\mu_i$}.}
\end{figure}

We now turn to the gluon mass equation; evidently, since its kernel is composed of \mbox{$\mu$-dependent} quantities, 
for each value of $\mu_i$ we will obtain a different solution, $m^2(q^2,\mu_i^2)$.
On the left panel of~\fig{fig:m_or} we show the corresponding solutions for the three renormalization points chosen.
The corresponding infrared saturation points, $m^2(0,\mu_i^2)= \Delta^{-1}(0,\mu_i^2)$,  
are given by \mbox{$m(0,\mu_1^2)=375$ MeV}, \mbox{$m(0,\mu_2^2)=412$ MeV}, and \mbox{$m(0,\mu_3^2)=434$ MeV}.
In addition, as anticipated, also the values of the arbitrary constant $C$ display a mild  $\mu$-dependence: 
$C(\mu_1)=9.2$, $C(\mu_2)=8.5$, and $C(\mu_3)=8.4$.


Now we are in the position to determine the behavior of the  RGI mass \mbox{${\overline m}^2(q^2)$}. To that end, 
we substitute into~\1eq{mRGIF} the  \mbox{$\mu$-dependent} results for  $m^2(q^2,\mu_i^2)$, $F(q^2,\mu_i^2)$ and $\alpha_s(\mu_i)$ obtained above. 
This is shown on the right panel of \fig{fig:m_or}, where we plot the quantity \mbox{${4\pi\overline m}^2(q^2)$} for 
the three values of $\mu_i$. As we can see, we have a nice agreement between the three curves in the range from $0$ to  \mbox{$0.05\,\mbox{GeV}^2$}. 
However, for higher values of $q^2$ they separate from the other, reaching the biggest discrepancy at \mbox{$q^2=7.5\,\mbox{GeV}^2$}, 
where the  percentage  error between  the (black) continuous and the (blue) dashed curves is around $64\%$. 

Evidently, the considerable deviations from the exact RG-invariance  
displayed in~\fig{fig:m_or} indicate that the form of the function 
$S$ employed in the  mass equation  needs to be improved.  
As we will see in the next section with some specific examples, 
such an improvement is indeed possible, and can be obtained 
by resorting to basic RGI arguments.

\section{\label{RGimp} RG improved versions of the kernel}

As has been established in \1eq{GRGI}, 
the quantity ${\cal G}$ 
must be such that, when multiplied by 
$g \Delta^{1/2}(q) \Delta^{1/2}(r) \Delta^{1/2}(p)$,
ought to give rise to an RGI combination. 
This information may be used to obtain some 
well-motivated Ans\"atze for $S$, which, in turn, may lead to 
solutions for $m^2(q^2)$ that are better behaved under the RG. 
In this section we will explore the explicit realization of this possibility. 
The study presented here is by no means exhaustive; it is simply indicative 
of how RG-improved versions of the gluon mass equation may be obtained in principle.

\subsection{Two simple models}

Specifically, let us set 
\be
S(q,k,k+q) = F(q) W(k,k+q),
\label{Sanz}
\ee
where, in accordance with the general properties of $S$, the $W$ is symmetric  under the exchange $k \leftrightarrow k+q$.
In addition, at tree level 
we must have $W^{(0)}=1$, so that, since $F^{(0)}(q)=1$,  
we get $S^{(0)}=1$, as required. 

If we now use the $S$ of \1eq{Sanz} to construct the lhs of \1eq{GRGI}, we have 
\bea
g \Delta^{1/2}(q) \Delta^{1/2}(k) \Delta^{1/2}(k+q) S &=& [g F(q)\Delta^{1/2}(q)] 
\left\{\Delta^{1/2}(k) \Delta^{1/2}(k+q) W(k,k+q)\right\} 
\nonumber\\
&=& d^{1/2}(q^2)\left\{\Delta^{1/2}(k) \Delta^{1/2}(k+q) W(k,k+q) \right\}. 
\eea
It is clear now that the presence of $F(q)$ facilitates the realization of the RGI combination, 
by providing the missing ingredient for the formation of $d^{1/2}(q^2)$; 
it is, therefore, an advantageous starting point.
The remaining structure must obviously come from $W$, which must convert the combination 
inside the curly bracket into another RGI quantity. 

In order to devise an approximate expression for the (dimensionless) $W$, let us first 
consider the one-loop expression of $\Delta^{-1}(k)=k^2 J(k)$, in the MOM scheme.
For the (dimensionless) $J(k^2)$ we have 
\be
J(k) = 1 + \frac{13}{6} t(k),
\ee
and so, 
\be
J^{1/2}(k) J^{1/2}(k+q) = 1 + \frac{13}{12}\, \bigg[ t(k) + t(k+q)\bigg] +  {\cal O}(\alpha^2_s)\,.
\ee
Thus, at order ${\cal O}(\alpha_s)$ the minimal necessary structure for $W$ is  
\be
W^{(1)} = 1+ \frac{13}{12}\, \bigg[ t(k) + t(k+q)\bigg] .
\label{W1}
\ee
Of course, this minimal form may be multiplied by a $\mu$-independent
function, which, at the given order, will provide the (unknown) rhs of \1eq{GRGI}.
Evidently, use of the minimal $W^{(1)}$ gives rise to a lhs equal to unity.

These observations motivate the study of two simple extensions of \1eq{W1}, 
where some additional structure has been added 
in order to model higher order effects or purely nonperturbative contributions. 

The cases we will consider are 
\be
W_1 =  1+ \frac{13}{12} \bigg[ t(k) + t(k+q)\bigg ] + c_1 \bigg[ t^2(k) + t^2(k+q)\bigg] + c_2 t(k)t(k+q),
\label{w1_model}
\ee
and
\be
W_2 = 1+ \frac{13}{12}\, \bigg[ t(k) + t(k+q)\bigg] + c.
\label{w2_model}
\ee
We next study the numerical implications of the above two possibilities. 

\subsection{Numerical analysis} 


\begin{figure}[!t]
\hspace{-1.5cm}\includegraphics[scale=1]{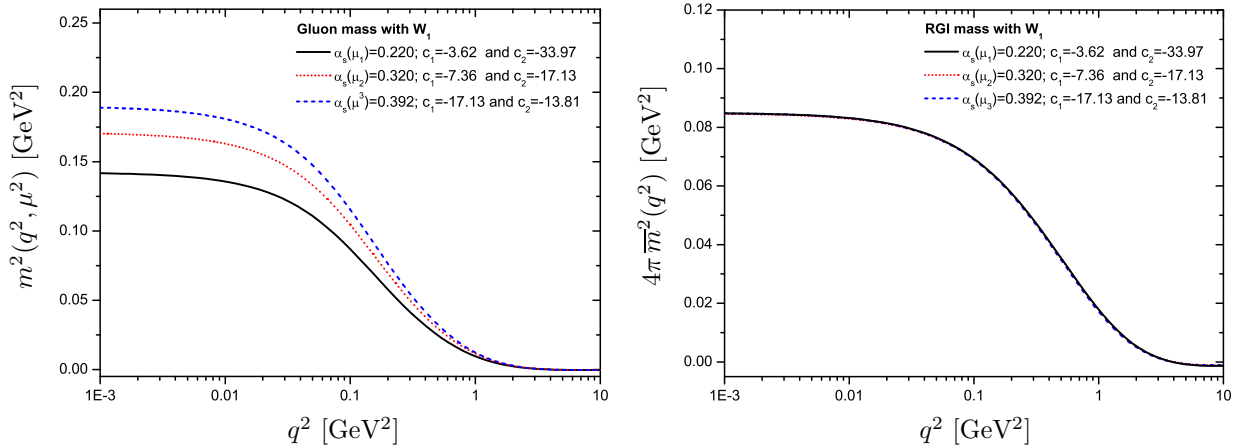}
\caption{\label{fig:case1} (color online)
The gluon dynamical mass  \mbox{$m^2(q^2,\mu^2)$} (left panel)
and the corresponding RGI mass \mbox{$4\pi{\overline m}^2(q^2)$} (right panel) obtained
using the model of~\1eq{w1_model} for the three values of $\mu_i$.}
\end{figure}


On the left panel of~\fig{fig:case1} we show the numerical solution for \mbox{$m^2(q^2,\mu_i^2)$} using the
model presented in~\1eq{w1_model}. In particular,  we choose, for the three different $\mu_i$, the parameters 
\mbox{$c_1(\mu_1)=-3.62$} and \mbox{$c_2(\mu_1)=-33.97$};  \mbox{$c_1(\mu_2)=-7.36$} and \mbox{$c_2(\mu_2)=-17.13$};  
\mbox{$c_1(\mu_3)=-7.42$} and \mbox{$c_2(\mu_3)=-13.81$}.

Although the 
general qualitative behavior of $m^2(q^2)$ appears rather similar
to that shown in~\fig{fig:m_or}, the 
RGI masses obtained from them 
show a definite improvement 
with respect to those of~\fig{fig:m_or}.  
Indeed, as one can clearly see  on the right panel of~\fig{fig:case1},
the three curves coincide within a wider range of momenta than in the previous case. 
More specifically, the less favorable region of momenta is around 
\mbox{$ q^2\approx 3.5\, \mbox{GeV}^2$}, where the relative error between the curves is smaller than $12\%$.
However, the downside of the form $W_1$ is that the appearance of a negative UV tail  for   $m^2(q^2)$  
(past~\mbox{ $ q^2\approx 3.5\,\mbox{GeV}^2$}) persists.       


\begin{figure}[!t]
\hspace{-1.5cm}\includegraphics[scale=1]{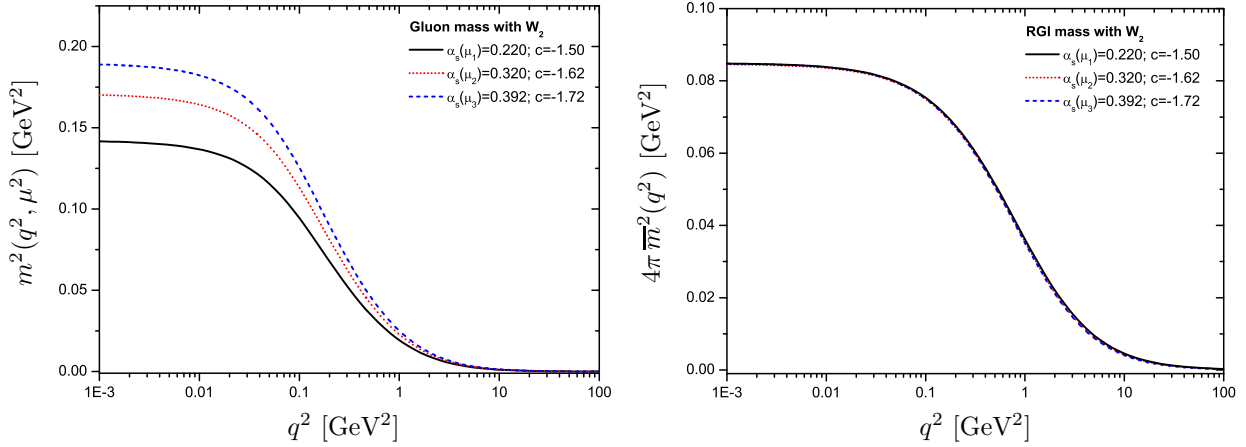}
\caption{\label{fig:case2} (color online) The numerical solution for \mbox{$m^2(q^2,\mu^2)$} (left panel) using 
the model of \1eq{w2_model} for the three renormalization scale $\mu_i$; the corresponding RGI mass \mbox{$4\pi{\overline m}^2(q^2)$} given by ~\1eq{mRGI} is shown on the right panel.}
\end{figure}

It is important to mention that the mass equation admits solutions for 
a variety of additional choices for $c_1(\mu_i)$ and  $c_2(\mu_i)$;   
however, the particular values quoted above are singled out because they yield 
\mbox{${\overline m}^2(q^2)$} that are as close to being perfectly RGI as possible.
In that sense, our scanning through the possible values of $c_1(\mu_i)$ and  $c_2(\mu_i)$ 
is by no means exhaustive, but only indicative of certain general trends in the type of solutions obtained. 

We next analyze the second model, where~\1eq{w2_model} is used in the kernel of the gluon mass equation. 
The results for this particular case are presented in~\fig{fig:case2}. On the left panel we plot \mbox{$m^2(q^2,\mu_i^2)$},
for the three values of $\mu_i$ chosen. 
The right panel shows the RGI quantity \mbox{$4\pi{\overline m}^2(q^2)$}; 
evidently, the results for the three $\mu_i$ practically collapse on a unique curve.
In fact, the less favorable point is located at \mbox{ $ q^2\approx 25\,\mbox{GeV}^2$}, where
the relative error is around $10\%$. In addition, the solutions obtained with   
$W_2$ remain positive and monotonically decreasing through the {\it entire range} of physical momenta.  
For the curves presented in ~\1eq{w2_model}, we have chosen $c=-1.50$ for $\mu_1$; $c=-1.62$ for $\mu_2$, and
$c=-1.72$  for $\mu_3$. 


\begin{figure}[!t]
\includegraphics[scale=1]{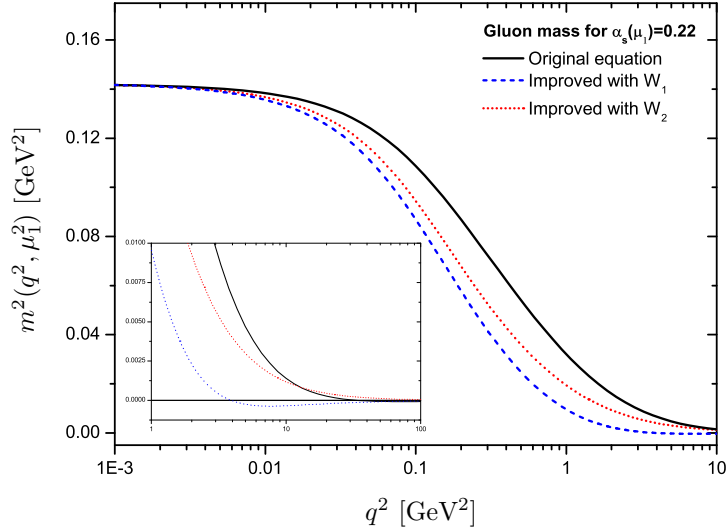}
\caption{\label{fig:comp_mass}(color online) Comparison of the numerical results obtained for  \mbox{$m^2(q^2,\mu^2_1)$} using
the original gluon mass equation (black continuous); the improved versions with $W_1$ of \1eq{w1_model}(blue dashed) and
$W_2$ of \1eq{w2_model} (red dotted). The inset shows a zoom in the UV tail of the same quantities.}
\end{figure}


In~\fig{fig:comp_mass} 
we compare the numerical solutions for \mbox{$m^2(q^2,\mu_1^2)$}  obtained from the three different models.
The solution of the original version of the gluon mass equation is represented
by the (black) continuous curve, while 
the solutions using ~\1eq{w1_model} and~\1eq{w2_model} are 
indicated by (blue) dashed and (red) dotted curves, respectively.
When the gluon propagator is renormalized at the 
point $\mu_1$, its saturation value in the deep IR is given by 
\be
\Delta^{-1}(0) = 0.14 \,{\rm GeV}^2 = m^2(0)\equiv m_0^2.
\ee
Therefore, the  three masses coincide at the origin, {\it i.e.}, \mbox{$m_0=375$ MeV}. However, 
in the intermediate region we clearly see differences in their momentum dependence. 
Notice that, in this particular region, the original equation produces a \mbox{$m^2(q^2)$} that falls off slower than the improved versions. 
On the other hand, the \mbox{$m^2(q^2)$} obtained with the $W_1$ decreases considerably faster than the other two cases. 
The UV tails of these solutions are shown separately in the insert; as already mentioned, 
only the one originating from ~\1eq{w2_model} stays strictly positive for all momenta.

\subsection{A physically motivated fit}

It turns out that 
the three different masses in~\fig{fig:comp_mass} may be fitted very accurately by a single, particularly simple 
function, namely 
\be
m^2(q^2) = \frac{m^2_0}{1 + \left(q^2/{\cal M}^2\right)^{1+p}} . 
\label{fit}
\ee 
The corresponding sets of optimal values,  $({\cal M},p)$, 
for the mass scale ${\cal M}$ and the exponent $p$ 
are as follows: 
\n{\it i} (557 MeV, 0.08) for the black continuous curve;
\n{\it ii} (381 MeV, 0.26)   for the blue dashed curve;
\n{\it iii}  (436 MeV, 0.15) for the red dotted curve. 
All fits have a reduced \mbox{$\chi^2 \approx 0.99$}. 
Note that ${\cal M}$ is just a 
dimensionful fitting parameter, not to be confused with the 
 characteristic QCD mass scale, $\Lambda$; 
in fact, within the MOM scheme that we use, and for $\alpha_s=0.22$, 
we have that $\Lambda_{\rm\s{MOM}} =280$ MeV~\cite{Boucaud:2008gn}.


\begin{figure}[!t]
\includegraphics[scale=1]{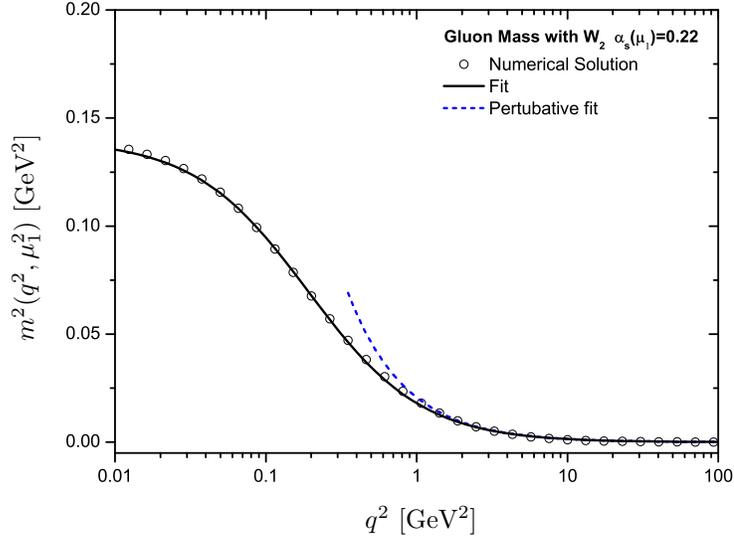}
\caption{\label{fig:fit}(color online) The numerical solution for \mbox{$m^2(q^2)$} obtained using the RG-improved Ansatz  
$W_2$ of \1eq{w2_model} (black circles). The (black) continuous curve represents the fit of~\1eq{fit}  while the (blue) dashed curve
is the asymptotic fit for the ultraviolet tail given by~\1eq{asfit}.}
\end{figure}


In order to appreciate the quality of the above fit, 
in~\fig{fig:fit} we superimpose the numerical solution for the RG-improved Ansatz  
$W_2$ when $\alpha(\mu_1)=0.22$ (black circles) and the fit of~\1eq{fit}(black continuous curve). 
Clearly, the coincidence between the two curves is striking. 

Let us now take a closer look at the asymptotic form of $m^2(q^2)$ for large $q^2$,  
 to be denoted by   $m^2_{\rm{\s {UV}}}(q^2)$.   
From \1eq{fit} it is clear that, for sufficiently large values of $q^2$, the 1  may be 
depreciated in the denominator of  \1eq{fit}, yielding 
\bea
m^2_{\rm{\s {UV}}}(q^2) = \frac{m^2_0 \Lambda^2}{q^2} \left(q^2/{\cal M}^2\right)^{-p}  .
\label{asfit}
\eea  
As is shown in ~\fig{fig:fit}, the onset of the asymptotic form 
\noeq{asfit} is clear already at momenta of the  order of a few GeV.

The particular asymptotic behavior described in \1eq{asfit}
corresponds precisely to the 
so-called ``power-law'' running  of the effective gluon mass, first conjectured in~\cite{Cornwall:1985bg}, 
and subsequently studied in~\cite{Lavelle:1991ve,Aguilar:2007ie}. 
Its main physical implication is that 
the condensates of dimension two do not contribute to the OPE expansion of $m^2_{\rm{\s {UV}}}(q^2)$,   
because otherwise  the corresponding running would have been logarithmic.
Then, in the absence of quarks, 
the lowest order condesates appearing in the OPE of the mass 
are those of dimension four, namely the (gauge-invariant) $\langle 0|\hspace{-0.125cm}:\hspace{-0.125cm}G_{\mu\nu}^{a}
G^{\mu\nu}_{a}\hspace{-0.125cm}:\hspace{-0.125cm}|0  \rangle$, and possibly the ghost condensate \mbox{$\langle 0|\hspace{-0.125cm}:\hspace{-0.125cm}{\overline c}^{a} \,\Box \, c^{a} \hspace{-0.125cm}:\hspace{-0.125cm}|0 \rangle$}~\cite{Lavelle:1988eg,Bagan:1989gt}.    
Since, on dimensional grounds, 
these condensates must be divided by $q^2$, 
one obtains (up to logarithms) the aforementioned power-law running for the mass. 

It remains to be seen if 
\1eq{asfit} is a fortuitous coincidence 
related to a particular Ansatz for the kernel (namely $W_2$),  
or if it really reflects an intrinsic feature of the gluon mass. 

\section{\label{conc}Discussion and Conclusions}
    
In this work we have presented a detailed study of the RG structure of the  integral equation that  controls the  dynamical evolution of the
gluon mass.   Specifically, we have shown that  the renormalization of
this  equation   can  be  carried   out  entirely  by  means   of  the
renormalization  constants  employed   in  the  standard  perturbative
treatment,  namely those  associated  with the  gluon  and ghost  wave
functions and the fundamental vertices of the theory.  In addition, by
making  explicit  use  of  the diagrammatic  equivalence  between  the
skeleton  expansion  of the  three  gluon vertex  in  the  SDE and  BS
formalisms,  the kernel  of the  gluon  mass equation  can be  written
exclusively in  terms of the  renormalized Green's functions,  with no
reference   to    any   cutoff-dependent   renormalization   constants
[see~\1eq{mren2}].

The  RG properties of the full mass equation 
are inevitably distorted when approximate expressions are used for its kernel.
The departure of the solutions from the correct RG behavior is quantitatively 
described in terms of the RGI gluon mass, \mbox{${\overline m}^2(q^2)$}, 
and can serve as a discriminant for the various Ans\"atze employed for the kernel. 
Using this criterion, 
we have established that the \mbox{${\overline m}^2(q^2)$} 
constructed using as input the solution $m^2(q^2)$
obtained from the original version of 
the mass equation~\cite{Binosi:2012sj}, 
deviates  considerably from the optimal RGI behavior (see~\fig{fig:m_or}).

Then, motivated by the RG properties that the kernel must satisfy, two 
new versions of the gluon mass equation were put forth [see ~\2eqs{w1_model}{w2_model}], 
which are expected to display an improved RG behavior.
Indeed, our numerical analysis reveals that the \mbox{${\overline m}^2(q^2)$} 
obtained from both RG-improved Ans\"atze capture more faithfully the RG properties of the exact equation.
Specifically, the deviations between the \mbox{${\overline m}^2(q^2)$} 
obtained for different $\mu$'s displays, in the less favorable regions, 
a relative error around $12\%$ and $10\%$, respectively. 
In addition, and contrary to the other two cases, the Ansatz of~\1eq{w2_model} presents a well-defined 
positive UV tail in all range of momenta. We therefore conclude that, overall, the  
best available functional form for the kernel is given by~\1eq{w2_model}. 

Interestingly enough, $W_2$ has a simpler structure than $W_1$, 
in the sense that it contains a single adjustable parameter instead of two, 
and yet it produces results that are in better compliance with the 
basic theoretical principles that we have considered. The reason for that may be related to the 
overall sign of the gluon mass equations, and the degree at which each Ansatz  
succeeds to effectively reverse it.
Specifically, as already mentioned after \1eq{tdef}, the negative sign on the rhs of \1eq{masseqnew}
must be compensated by negative contributions coming from the kernel. 
In the case of  $W_2$ this is accomplished directly, and in a rather elementary way,  
because the parameter $c$ is simply chosen such that $1+c$ becomes sufficiently negative. Instead, 
$W_1$ performs the same task in a less efficient way, which may be reflected in the  
slightly enhanced departure of the resulting mass from the perfect RG-invariance, 
and the change of its sign in the deep UV.  

It is clear that a more rigorous  determination of the kernel is required, in order to 
further substantiate our analysis. 
It is worth pointing out that, in this 
effort, one may want to keep open the possibility of working with the lhs of \1eq{airen}, rather than its rhs.
Indeed, whereas for the formal demonstration presented in Section~\ref{RGprop}
the rhs of \1eq{airen} seems to be advantageous, because it is free of the renormalization constants $Z$, 
for the actual computation of ${\cal G}^{\mu\alpha\beta}$ the lhs may turn out to be easier to handle. 
Of course, in order to make progress with the lhs, in addition to obtaining a  better approximation for the 
quantity $Y(k)$,  
one ought to provide appropriate expressions for the renormalization constants $Z$. 
These tasks are  technically particularly subtle and laborious, because they require, among other things, a 
tight control on the structure of the various fully-dressed vertices of the theory. In fact, 
the multiplicative renormalizability and the 
correct cancellation of overlapping divergences depends crucially on the detailed knowledge of the 
transverse (automatically conserved) part of the corresponding vertex 
(in our case of the three-gluon vertex), which forces one to go beyond the usual 
gauge-technique inspired Ans\"atze for the vertex in question~\cite{Kizilersu:2009kg}.  
These difficulties have been exemplified, and only partially circumvented, in the 
studies of the gap equation that controls the chiral symmetry breaking and the dynamical generation of a 
constituent quark mass~\cite{Roberts:1994dr,Curtis:1993py,Kizilersu:2009kg,Aguilar:2010cn}. 
We hope to make progress on some of these issues in future works.

\acknowledgments 

The research of J.~P. is supported by the Spanish MEYC under 
grant FPA2011-23596. The research  of  A.~C.~A  is supported by the 
National Council for Scientific and Technological Development - CNPq
under the grant 306537/2012-5 and project 473260/2012-3,
and by S\~ao Paulo Research Foundation - FAPESP through the project 2012/15643-1.

\end{document}